\documentclass[%
 reprint,
nofootinbib,
 amsmath,amssymb,
 aps,
]{revtex4-2}

\usepackage{graphicx}
\usepackage{dcolumn}
\usepackage{bm}

\usepackage{color,epstopdf,graphics,mathtools,physics}

\newcommand{\Mp}{M_{\rm Pl}}

\begin{document}

\preprint{APS/123-QED}

\title{Cosmological memory effect in scalar-tensor theories}

\author{Mohammad Ali Gorji}
\email{gorji@yukawa.kyoto-u.ac.jp}
\affiliation{%
Center for Gravitational Physics and Quantum Information, \\ 
Yukawa Institute for Theoretical Physics, Kyoto University, Kyoto 606-8502, Japan}

\author{Taisuke Matsuda}%
 \email{taisuke.matsuda@yukawa.kyoto-u.ac.jp}
\affiliation{%
Center for Gravitational Physics and Quantum Information, \\ 
Yukawa Institute for Theoretical Physics, Kyoto University, Kyoto 606-8502, Japan}

\author{Shinji Mukohyama}
\email{shinji.mukohyama@yukawa.kyoto-u.ac.jp}
\affiliation{Center for Gravitational Physics and Quantum Information, \\ 
	Yukawa Institute for Theoretical Physics, Kyoto University, Kyoto 606-8502, Japan
}%
\affiliation{
Kavli Institute for the Physics and Mathematics of the Universe (WPI), \\
The University of Tokyo, Chiba 277-8583, Japan
}%

\date{\today}

\begin{abstract}
The cosmological memory effect is a permanent change in the relative separation of test particles located in a FLRW spacetime due to the passage of gravitational waves. In the case of a spatially flat FLRW spacetime filled with a perfect fluid in general relativity, it is known that only tensor perturbations contribute to the memory effect while scalar and vector perturbations do not. In this paper, we show that in the context of scalar-tensor theories, the scalar perturbations associated to the scalar graviton contribute to the memory effect as well. We find that, depending on the mass and coupling, the influence of cosmic expansion on the memory effect due to the scalar perturbations can be either stronger or weaker than the one induced by the tensor perturbations. As a byproduct, in an appendix, we develop a general framework which can be used to study coupled wave equations in any curved spacetime region which admits a foliation by time slices.
\end{abstract}

\maketitle


\section{Introduction}

The recent detection of gravitational waves (GWs) has opened up a new window to probe different aspects of gravitational interaction that otherwise are impossible to be explored \cite{LIGOScientific:2017vwq,LIGOScientific:2017zic}. As one of the observable byproduct of the GWs, the so-called memory effect, which is a permanent and sudden change of the relative distance between two observers through the passage of GWs, has received a lots of attention in recent years. This effect was first indicated by Zel'dovich and Polnarev \cite{Zeldovich:1974gvh} and later completed by Christodoulou by taking into account nonlinear effects \cite{Christodoulou:1991cr} (see also \cite{Wiseman:1991ss,Thorne:1992sdb}).

In combination with the soft theorem and the asymptotic symmetry, the memory effect constitutes the so-called ``infrared triangle'' \cite{Strominger:2014pwa}. The soft theorem was first discovered in the context of QED \cite{Bloch:1937pw} and a few years later generalized by Weinberg to theories involving particles with arbitrary spins including gravitons \cite{Weinberg:1965nx}. In QED, it is implemented to cancel infrared divergences to preserve the consistency of the field theory \cite{Peskin:1995ev}. On the other hand, the known asymptotic symmetry for gravity is the so-called Bondi, van der Burg, Metzner and Sachs (BMS) symmetry which is the symmetry group of diffeomorphism transformations that do not break asymptotic flatness at conformal infinities \cite{Bondi:1962px,Sachs:1962wk}. The three corners of the infrared triangle are expected to be equivalent to each other: i) equivalence of BMS symmetry with the soft graviton theorem can be shown by Ward-Takahashi identity \cite{He:2014laa}, ii) the permanent change of the distance between observers caused by the memory effect can be realized through a step functional change of the spacetime metric and this change is indeed the same as the one generated by the BMS transformations \cite{Strominger:2014pwa}, iii) the change in an asymptotic metric due to the memory effect has the same form as the coefficient of the scattering amplitude due to the addition of soft gravitons in the soft theorem \cite{He:2014laa}. It is worth mentioning that these three subjects have been developed independently and these equivalences are quite nontrivial. Moreover, this triangle equivalence is not restricted to the gravitational theory and similar equivalences also show up in some gauge theories \cite{Kapec:2015ena,Strominger:2013lka}. Therefore, infrared triangle may have a deeper origin arising from the infrared consistency of a theory under consideration.

In order to better understand the infrared structure of gravity, it is then quite important to further study each corner of the infrared triangle. The memory effect, which is the subject of this paper, is originally found in Minkowski spacetime and later studied in asymptotically flat spacetime. In these cases, the radiation part of the gravitational field, which includes the memory effect, can be distinguished from the other tidal gravitational effects by looking at the fall off of the gravitational field near the spatial and null infinities. In the case of the cosmological spacetime, which is not asymptotically flat, characterizing the memory effect is more subtle. Recently, this issue was studied by different groups with different approaches \cite{Bieri:2015jwa,Chu:2015yua,Kehagias:2016zry,Chu:2016qxp,Bieri:2017vni,Bonga:2017dlx,Jokela:2022rhk} and among them, we will focus on the approach adopted by Tolish and Wald \cite{Tolish:2016ggo}. Indeed, using the fact that the spatially flat Friedmann-Lema\^{i}tre-Robertson-Walker (FLRW) spacetime is conformally flat, they have developed a general setup to study memory effect which is applicable as far as an idealized particle-like source for GWs is considered. Apart from the fact that we need to clarify the notion of memory in a cosmological background to study the universe, their setup provides a framework to better understand the infrared regime of gravity at cosmological scales. Their analysis of cosmological memory effect is based on general relativity and it will be interesting to explore what happens in modified theories of gravity. In this paper, we focus on scalar-tensor theories. According to the results of Ref. \cite{Tolish:2016ggo} for the linear perturbations, only tensor perturbations contribute to the cosmological memory effect in general relativity. We will show that, in scalar-tensor theories, scalar perturbations also contribute to the cosmological memory. 

The rest of the paper is organized as follows. In Sec. \ref{sec-STT}, we present our scalar-tensor model which is coupled to a perfect fluid and particle-like sources, as a source of GWs. We then study the background equations and linear perturbations around a spatially flat FLRW spacetime. In Sec. \ref{sec-retarded-GF}, we focus on the scalar perturbations and we find the direct part of the corresponding retarded Green's function. Using this result, in Sec. \ref{sec-memory-effect}, we show that the scalar perturbations contribute to the cosmological memory effect in scalar-tensor theories. Sec. \ref{sec-summary} is devoted to the summary of the paper. Moreover, we present our model in the Jordan and the Einstein frames in Appendix \ref{app-EJ-frames} and we present relation between the energy-momentum tensors in different frames in Appendix \ref{app-EJ-EMT}. In Appendix \ref{app-MS-marices}, we show explicit forms of the mass and source matrices for the sake of completeness. In Appendix \ref{app-longitudinal-gauge}, we present scalar perturbations in terms of the gauge-invariant counterpart of the scalar field perturbation to make the minimal coupling and constant scalar background limit of the theory manifest. Finally, in Appendix \ref{app-Green}, we develop a general framework which can be used to study coupled wave equations in any curved spacetime region which admits a foliation by time slices. The cosmological spacetime, which we deal with in this paper, can be considered as a special subset.

\section{The scalar-tensor theory}\label{sec-STT}

We consider a scalar-tensor theory with a linear kinetic term and without higher derivative terms. The action of the system in the Jordan frame, in which matter fields directly couple to the metric, is given by
\begin{align}
S_{\rm J}[{\tilde g},\phi,\psi] 
&= \int d^4x\sqrt{-{\tilde g}} \bigg[ \frac{\Mp^2}{2} F(\phi) {\tilde R}
\nonumber \\ \label{action-J}
&- \frac{1}{2} {\tilde K}(\phi) {\tilde g}^{\mu\nu}\partial_{\mu}\phi \partial_{\nu}\phi - {\tilde V}(\phi)
\bigg] + S_{\rm m}[{\tilde g},\psi] \,,
\end{align}
where $\Mp=(8\pi{G})^{-1/2}$ is the reduced Planck mass, ${\tilde R}$ is the Ricci scalar in the Jordan frame, $F$, ${\tilde K}$ and ${\tilde V}$ are functions of the scalar field, and $S_{\rm m}$ is the matter action in which $\psi$ collectively represents all matter fields and particles which are present in the system under consideration. The gravitational part of the action \eqref{action-J} is constructed out of the metric in the Jordan frame ${\tilde g}_{\mu\nu}$ and scalar field $\phi$. Performing the conformal transformation
\begin{equation}\label{CT}
{\tilde g}_{\mu\nu} = F(\phi)^{-1} g_{\mu\nu} \,,
\end{equation}
where $g_{\mu\nu}$ is the metric in the Einstein frame, the action in the Einstein frame takes the form (see Appendix \ref{app-EJ-frames})
\begin{align}\label{action-E}
S_{\rm E}[{g}, \varphi,\psi] &= \int d^4x\sqrt{-{ g}} \bigg[
\frac{\Mp^2}{2} { R} 
\nonumber \\
&
- \frac{1}{2} { g}^{\mu\nu}\partial_{\mu}\varphi \partial_{\nu}\varphi - { V}(\varphi)
\bigg]
+ S_{\rm m}[\tilde{g}, \psi]  \,,
\end{align}
where we have defined
\begin{align}
\varphi \equiv \int \sqrt{K(\phi)} \, d\phi , \hspace{.2cm}
K \equiv \frac{{\tilde K}}{F} + \frac{3}{2} \Mp^2 \left( \frac{F_{,\phi}}{F} \right)^2 , 
\hspace{.2cm}
V \equiv \frac{{\tilde V}}{F^2} ,
\end{align}
and it is understood that $\tilde{g}$ in \eqref{action-E} is given by \eqref{CT} and that $\phi$ is considered as a function of $\varphi$. 

Comparing the equivalent actions \eqref{action-J} and \eqref{action-E}, we find that the form of the action for $g_{\mu\nu}$ is different from that for $\tilde{g}_{\mu\nu}$. This difference is compensated by the fact that the coupling to the matter sources is different: the matter couples only to ${\tilde g}_{\mu\nu}$ in the Jordan frame while it couples to both $g_{\mu\nu}$ and $\varphi$ in the Einstein frame. As it is well-known, this is the reason why we consider the scalar-tensor theories as modified gravity theories, despite the fact that $g_{\mu\nu}$ is described by the Einstein-Hilbert action as in general relativity. 

As it is easier, we perform all calculations in the Einstein frame and translate only the final results in terms of the Jordan frame quantities. The details of the transformation between the Jordan frame and the Einstein frame are presented in Appendix \ref{app-EJ-frames}. From now on (throughout this and next sections), all calculations are presented in the Einstein frame. The Einstein equations can be deduced by taking the variation of the action \eqref{action-E} with respect to the metric $g^{\mu\nu}$
\begin{eqnarray}\label{EE-ST}
\Mp^2 G_{\mu\nu} = T^{(\varphi)}_{\mu\nu} + T^{({\rm m})}_{\mu\nu} \,; \hspace{.5cm}
T^{({\rm m})}_{\mu\nu} = \frac{-2}{\sqrt{- g}} \frac{\delta{S}_{\rm m}}{\delta{ g}^{\mu\nu}} \,,
\end{eqnarray}
where  $G_{\mu\nu}$ is the Einstein tensor and $T^{({\rm m})}_{\mu\nu}$ is the energy-momentum tensor of the matter (see Appendix~\ref{app-EJ-EMT}) while energy-momentum tensor of the scalar field is given by
\begin{eqnarray}\label{EMT-phi}
T^{(\varphi)}_{\mu\nu} = \partial_{\mu} \varphi \partial_{\nu} \varphi
- g_{\mu\nu} \left[
\frac{1}{2} g^{\alpha\beta}\partial_{\alpha}\varphi \partial_{\beta}\varphi + V(\varphi) 
\right] \,.
\end{eqnarray}
Taking variation of the action \eqref{action-E} with respect to the scalar field $\varphi$, we find
\begin{eqnarray}\label{EOM-varphi}
\Box{\varphi} - V_{,\varphi} = \frac{F_{,\varphi}}{2F} \, T^{({\rm m})} \,; \hspace{.5cm}
T^{({\rm m})} = g^{\alpha\beta} T^{({\rm m})}_{\alpha\beta} \,.
\end{eqnarray}

We consider two types of sources for the matter sector: perfect fluid which is responsible for the energy density of the universe and particle-like sources as idealized sources for GWs production. The matter energy-momentum tensor then can be separated to two parts
\begin{eqnarray}
T^{({\rm m})}_{\mu\nu} =  { T}^{({\rm F})}_{\mu\nu} + { T}^{({\rm P})}_{\mu\nu} \,.
\end{eqnarray}

The conservation equations for the matter [see Eq. \eqref{CoE}] imply
\begin{eqnarray} \label{EOM-COM-F}
\nabla_{\mu} { T}^{({\rm F})\mu}{}_{\nu}
= - \frac{F_{,\varphi}}{2F} \, { T}^{({\rm F})} \nabla_{\nu}\varphi \,,
\end{eqnarray}
\begin{eqnarray} \label{EOM-COM-P}
\nabla_{\mu} { T}^{({\rm P})\mu}{}_{\nu}
= - \frac{F_{,\varphi}}{2F} \, { T}^{({\rm P})} \nabla_{\nu}\varphi \,.
\end{eqnarray}

The energy-momentum tensor for the perfect fluid is given by
\begin{equation}\label{EMT-F}
{ T}^{({\rm F})}_{\mu\nu} = (\rho+p) { U}_\mu { U}_\nu + p \, { g}_{\mu\nu} \,,
\end{equation}
where $\rho$ and $p$ are the energy density and pressure in the Einstein frame while $U_\mu$ is the four-velocity normalized with respect to the Einstein frame metric $g^{\mu\nu}{ U}_\mu{ U}_\nu = -1$.

For the particle energy-momentum tensor we have
\begin{align}
{ T}^{({\rm P})}_{\mu\nu} &= 
\sum_{l,{\rm in}} { T}^{(M,l)}_{\mu\nu} + \sum_{n,{\rm in}} { T}^{(N,n)}_{\mu\nu}
\nonumber \\ \label{EMT-P-J}
&+ \sum_{l',{\rm out}} { T}^{(M,l')}_{\mu\nu} + \sum_{n',{\rm out}} { T}^{(N,n')}_{\mu\nu} \,,   
\end{align}
in which the upper indices $M$ and $N$ indicate massive and massless particles respectively with
\begin{eqnarray}\label{EMT-M}
T^{(M,{\rm in})}_{\mu\nu} &=& 
m u_\mu u_\nu \delta^{(3)}({\bf x}-{\bf z}(t)) \frac{1}{\sqrt{-{ g}}} \frac{d\tau}{dt} \Theta(-t) \,,
\\ 
\label{EMT-N}
T^{(N,{\rm in})}_{\mu\nu} &=& 
k_\mu k_\nu \delta^{(3)}({\bf x}-{\bf y}(t)) \frac{1}{\sqrt{-{ g}}} \frac{ed\lambda}{dt} \Theta(-t) \,,
\end{eqnarray}
where $u_{\mu}= \frac{d x^{\mu}}{d\tau} $ is the four-velocity of the massive particles and $k^{\mu}=\frac{d x^{\mu}}{ed\lambda}$ is the four-momentum of the massless particles. Notice that $\tau$ is the proper time while $t$ is the local time coordinate for the particles. The label ``in'' shows that the particles come from the past lightcone for $t<0$ while the label ``out'' corresponds to the particles in the future lightcone with $t>0$. Therefore, the energy-momentum tensor for the outgoing massive and massless particles have the same forms as Eqs. \eqref{EMT-M} and \eqref{EMT-N} with the replacement of $\Theta(-t)$ with $\Theta(t)$.

\subsection{Background equations}

For the background configuration of the gravity part of the system \eqref{action-E}, we consider a spatially flat FLRW metric and a homogeneous profile of the scalar field
\begin{align}
ds^2 &= {\bar g}_{\mu\nu} dx^\mu dx^\nu 
\nonumber \\ \label{BG-gravity}
&= - \bar{N}(\tau)^2 d\tau^2+ a(\tau)^2 \delta_{ij} dx^i dx^j \,, 
\hspace{.2cm} 
{\bar \varphi} = {\bar \varphi}(\tau) \,,
\end{align}
where $\bar{N}$ and $a$ are the lapse function and the scale factor. All quantities with bar denote the corresponding background values which only depend on $\tau$.

For the perfect fluid in the matter sector, we consider the homogeneous and isotropic configuration
\begin{equation}\label{BG-PF}
{\bar T}^{({\rm F})}_{\mu\nu} 
= ({\bar \rho}+\bar{p}) \, {\bar U}_\mu {\bar U}_\nu + \bar{p} \, {\bar g}_{\mu\nu} \,;
\hspace{.2cm}
{\bar U}_{0} = - \bar{N} \,,
\hspace{.2cm} 
{\bar U}_{i} = 0 \,.
\end{equation}
Note that the particle-like sources in the matter sector do not have non-vanishing background values and will show up only at the level of perturbations.

The Einstein Eqs. \eqref{EE-ST} for the background configuration give the Friedmann equations
\begin{eqnarray}
3 \Mp^2 H^2 &=& {\bar \rho} + {\bar \rho}_{\varphi} \,,
\\ \label{Hdot}
- \Mp^2 \left( 2 \dot{H}+ 3 H^2 \right) &=& {\bar p} + {\bar p}_{\varphi} \,,
\end{eqnarray}
where a dot denotes derivative with respect to the cosmic time $d/(\bar{N}d\tau)$ and $H=\dot{a}/a$ is the Hubble expansion rate. The energy density and pressure of the scalar field at the background level are
\begin{eqnarray}
{\bar \rho}_{\varphi} = \frac{1}{2} \dot{\bar \varphi}^2 + {\bar V} \,,
\hspace{.5cm}
{\bar p}_{\varphi} = \frac{1}{2} \dot{\bar \varphi}^2 - {\bar V} \,.
\end{eqnarray}

The equation of motion for the scalar field \eqref{EOM-varphi} gives
\begin{equation}\label{KG-BG}
\ddot{\bar \varphi} + 3 H \dot{\bar \varphi} + \bar{V}_{,\varphi} = 
\frac{\bar{F}_{,\varphi}}{2\bar{F}} \big( 1 - 3 w \big) {\bar \rho} \,,
\end{equation}
while the conservation equation for the perfect fluid \eqref{EOM-COM-F} implies
\begin{equation}
\dot{\bar\rho} + 3 H ( 1 + w ) {\bar \rho}
= - \frac{\bar{F}_{,\varphi}}{2\bar{F}} \big( 1 - 3 w \big) {\bar \rho} \dot{\bar \varphi} \,,
\end{equation}
where we have defined the equation of state parameter $w \equiv {{\bar p}}/{\bar \rho}$. The conservation equation for the particle-like sources \eqref{EOM-COM-P} trivially holds at the background level.

\subsection{Linear perturbations}

For the perturbations in the gravity sector, we consider\footnote{We have adopted the notation $A_{(i}B_{j)}=(A_iB_j+A_jB_i)/2$.}
\begin{eqnarray}\label{Pert-gravity}
&&\delta\varphi \,,
\hspace{1cm}
\delta{g}_{00} = - 2 \bar{N}^2 \alpha \,, 
\hspace{1cm}
\delta{g}_{0i} = \bar{N} \left( \partial_i \beta + \beta_i \right) \,, 
 \\ \nonumber
&&
\delta{g}_{ij} = a^2 \left[
2 \psi \delta_{ij} + 2 \left( \partial_i \partial_j - \frac{1}{3} \delta_{ij} \partial^2 \right) E + 2 \partial_{(i}C_{j)} + h_{ij}
\right] \,,
\end{eqnarray}
where $\{\alpha,\beta,\psi,E,\delta\varphi\}$ are scalar perturbations, $\{\beta_i,C_i\}$ are vector perturbations which are divergence-free $\partial^i\beta_i = 0 = \partial^iC_i$, and $h_{ij}$ are tensor perturbations which satisfy the traceless and transverse conditions $h^i{}_{i} = 0 = \partial^i h_{ij}$.

As we already mentioned, the particle energy-momentum tensor does not contribute to the background and it starts to contribute at the level of perturbations. Thus, we decompose it similarly to the metric perturbations as follows
\begin{align}
T^{(P)}_{00} &= - 2 \bar{N}^2 \alpha^{(P)} \,, 
\hspace{1cm}
T^{(P)} _{0i} = \bar{N} \left( \partial_i \beta^{(P)}  + \beta^{(P)}_i \right) \,, 
\nonumber \\
T^{(P)}_{ij} &= a^2 \bigg[ 2 \psi^{(P)} \delta_{ij} + 2 \left( \partial_i \partial_j - \frac{1}{3} \delta_{ij} \partial^2 \right) E^{(P)}  
\nonumber \\ \label{Pert-particle} 
& + 2 \partial_{(i}C_{j)}^{(P)} + {\mathcal T}_{ij}
\bigg] \,,
\end{align}
where $\{\alpha^{(P)},\beta^{(P)},\psi^{(P)},E^{(P)}\}$ are scalar perturbations, $\{\beta^{(P)}_i,C^{(P)}_i\}$ are divergence-free  vector perturbations, and ${\mathcal T}_{ij}$ are traceless and transverse tensor perturbations. 

For the perfect fluid, we define perturbations in the energy density and four-velocity as follows
\begin{equation}
\delta T^0{}_0 = - \delta\rho \,, \hspace{1cm} \delta{U}_i = - \partial_i U + U^T_i \,, 
\end{equation}
where $(\delta\rho,U)$ are scalar perturbations while $U^T_i$ are divergence-free vector perturbations. Perturbations in the pressure and temporal component of the four-velocity are not independent quantities
\begin{eqnarray}
\delta{p} = c_s^2 \, \delta\rho \,, \hspace{.5cm} \delta{U}_0 = - \bar{N} \alpha \,, \hspace{.5cm}
c_s^2 \equiv \frac{d{\bar p}}{d{\bar \rho}} \,,
\end{eqnarray}
where $c_s$ is the speed of sound for the scalar perturbations. 

Note that the vector and tensor perturbations are not affected by the scalar field $\varphi$ and, therefore, the results will be completely the same as those already studied in the context of general relativity \cite{Tolish:2016ggo}: the vector perturbations do not contribute to the memory effect while the tensor perturbations do contribute. Then, we do not consider the vector perturbations. We focus on the scalar perturbations while we briefly present the results for the tensor perturbations. 

The tensor perturbations are gauge-invariant and all perturbations in the particle energy-momentum tensor are gauge-invariant as well since the particle-like sources do not contribute to the background. The linearized Einstein Eqs. \eqref{EE-ST} for the tensor perturbations $h_{ij}$ and ${\cal T}_{ij}$ give
\begin{equation}\label{EoM-T}
\frac{\Mp^2}{2} \left( \ddot{h}_{ij} + 3 H \dot{h}_{ij} - \frac{1}{a^2} \partial^2 h_{ij} \right) = {\cal T}_{ij} \,.
\end{equation}

As the scalar perturbations are not gauge-invariant, we introduce the following gauge-invariant scalar perturbation in the gravity sector
\begin{eqnarray}\label{zeta-def-0}
\zeta \equiv \psi - \frac{H}{\dot{\bar\varphi}} \delta\varphi \,,
\end{eqnarray}
and we then work in the unitary gauge
\begin{align}\label{unitary-gauge}
\delta\varphi = 0 \,, \qquad E = 0 \,, \qquad \psi = \zeta \,.
\end{align}
In order to simplify the calculations, we perform the following transformation from $\delta\rho$ to $\delta\chi$ in the fluid sector\footnote{The transformation \eqref{Trans} is not a point transformation as it includes time derivative of the new field $\delta\chi$. One way to perform it is to implement Hamiltonian formalism where \eqref{Trans} can be considered as a canonical transformation. There is, however, a simpler way to perform \eqref{Trans} at the level of Lagrangian which is explained in Appendix B of Ref. \cite{DeFelice:2015moy}.}
\begin{align}\label{Trans}
\frac{\delta\rho}{\bar{\rho} } =\frac{1+w}{ c_s^2} 
\left( \frac{\dot{\delta\chi}}{v} - \alpha \right) \,,
\end{align}
where the background quantity $v$ is a solution of the following first-order differential equation\footnote{As it is known, scalar field models without higher derivative terms can be modeled into a perfect fluid. If one considers a shift-symmetric scalar field $\chi$ instead of the perfect fluid and performs the usual background/perturbation decomposition $\chi=\bar{\chi}+\delta\chi$, $\delta\chi$ coincides with $\delta\chi$ in Eq. \eqref{Trans} and Eq. \eqref{chi-BG} is the background equation for $\bar{\chi}$ with $v=\dot{\bar{\chi}}$.}
\begin{align}\label{chi-BG}
\dot{v} + 3 H c_s^2 v = \frac{\dot{\bar{\varphi }}\bar{F}_{\varphi }}{2 \bar{F}} \left(3 c_s^2-1\right) \,.
\end{align}
For an explicitly given background configuration, we can solve the above equation to find an explicit form of $v$. 

The linearized Einstein Eqs. \eqref{EE-ST} for the scalar perturbations $\zeta, \delta\chi, \alpha, \beta$ then give
\begin{align}
&2 \Mp^2 \left[ 3 H \big( \dot{\zeta} - \alpha H \big) 
- \frac{1}{a^2} \partial^2 \left( \zeta + H \beta \right) \right] 
\nonumber \\ \label{EEp-00-u}
&= 
- \alpha \dot{\bar{\varphi }}^2
+ \frac{1+w}{c_s^2} \bar{\rho }
\left( \frac{\dot{\delta \chi }}{v} - \alpha \right)
- 2 \alpha^{(P)} \,,
\end{align}
\begin{align}\label{EEp-0i-u}
2 \Mp^2 \partial_i \left(\dot{\zeta } - H\alpha \right) = - \partial_i \left[
\frac{\delta \chi}{v} (1+w) \bar{\rho } 
+ \beta^{(P)} \right] \,,
\end{align}
\begin{align}
&\frac{1}{a^2} \Mp^2 \partial^i \partial_j \left( \zeta + \alpha + \dot{\beta} + H \beta \right)
\nonumber \\ \label{EEp-ij-u}
&= - 2 \partial^i \partial_j E^{(P)} \,, \hspace{1cm} \mbox{for  }\quad i \neq j\,,
\end{align}
\begin{align}\label{EEp-ii-u}
&2 \Mp^2 
\bigg[ \ddot{\zeta} + H \left( 3\dot{\zeta }-\dot{\alpha } \right) 
- \left( 2\dot{H} + 3 H^2 \right) \alpha
\nonumber \\
&- \frac{1}{3a^2} \partial^2\left( \zeta + \alpha + \dot{\beta} + H \beta \right)
\bigg]
\nonumber \\
& 
= \alpha \dot{\bar{\varphi }}^2 
- (1+w) \bar{\rho } \left( \frac{\dot{\delta \chi }}{v} - \alpha \right)
- 2 \psi^{(P)}
\,, \hspace{1cm} \mbox{for}\quad i = j
\end{align}

Linearizing the temporal component of Eq. \eqref{EOM-COM-F}, we find
\begin{align}\label{COMF-p-t-u}
&
\frac{\ddot{\delta \chi} }{v}
+\frac{3 H }{2} \left[ \left(1-c_s^2\right) \frac{\dot{\delta \chi }}{v} 
+ (3 c_s^2 - 1) \alpha
\right]
\nonumber \\ 
&
+ 3 c_s^2 \dot{\zeta } - \dot{\alpha }
- \frac{ c_s^2 }{a^2} \partial^2 \left( \frac{\delta\chi}{v} + \beta \right)
\nonumber \\ 
&
= -\frac{\dot{\bar{\varphi }} \bar{F}_{\varphi } }{4 \bar{F}}
\left(3c_s^2-1\right)
\left(\frac{\dot{\delta \chi } }{v} -3 \alpha\right) \,,
\end{align}
where we have used Eq. \eqref{chi-BG} to remove $\dot{v}$. The spatial components of Eq. \eqref{EOM-COM-F} are automatically satisfied after substituting $\dot{v}$ from Eq. \eqref{chi-BG}. The linearized equation for the scalar field \eqref{EOM-varphi} also gives
\begin{align}\label{KGp-u}
&
3 \dot{\zeta} - \dot{\alpha }
+ \frac{2 \bar{V}_{\varphi }}{\dot{\bar{\varphi }}} \, \alpha 
- \frac{1}{a^2} \partial^2 \beta
\nonumber \\
&= 
- \frac{\bar{F}_{\varphi }}{\dot{\bar{\varphi }} \bar{F}} 
\Bigg\{
\alpha^{(P)} +3 \psi^{(P)}
\nonumber \\
&
+ (1+w) \frac{\bar{\rho } }{ 2c_s^2}
\bigg[ \left(1 - \frac{5-3 w}{1+w} c_s^2\right) \, \alpha
- \left(1-3 c_s^2\right) \, \frac{\dot{\delta \chi }}{v} \bigg]
\Bigg\}
\,,
\end{align}
where we have used \eqref{KG-BG} to remove $\ddot{\bar{\varphi}}$. Finally, linearizing equations for the conservation of particle energy-momentum tensor \eqref{EOM-COM-P}, we find 
\begin{align}
&\dot{\alpha}^{(P)}
+ 3 H \left(\alpha^{(P)}-\psi^{(P)} \right) + \frac{1}{2a^2} \partial^2\beta^{(P)}
\nonumber \\ \label{COMP-p-t-u}
&
=
- \frac{\dot{\bar{\varphi }}\bar{F}_{\varphi }}{2\bar{F}} 
\left( \alpha^{(P)} + 3 \psi^{(P)}\right) \,,
\\
\label{COMP-p-i-u}
& \dot{\beta}^{(P)} + 3H \beta^{(P)} - 2 \psi^{(P)} - \frac{4}{3} \partial^2 E^{(P)} = 0 \,.
\end{align}

Now, our task is to remove the non-dynamical fields $\alpha$ and $\beta$. First, we find $\alpha$ and $\beta$ from Eqs. \eqref{EEp-00-u} and \eqref{EEp-0i-u}. Using these results in Eqs. \eqref{EEp-ij-u} and \eqref{KGp-u} we find solutions for $\dot{\alpha}$ and $\dot{\beta}$. Substituting $\dot{\alpha}$, $\dot{\beta}$, $\alpha$, $\beta$ in Eqs. \eqref{EEp-ii-u} and \eqref{COMF-p-t-u} we find
\begin{align}\label{Eq-Matrix-ND}
&{\boldsymbol{\mathcal L}}. \boldsymbol{\xi} = - 4\pi \boldsymbol{\mu}^{(P)} \,;
\\ \nonumber
&{\boldsymbol{\mathcal L}} \equiv - 
\left[ \boldsymbol{1} \frac{d}{{\bar N}d\tau} 
\left( \frac{d}{{\bar N}d\tau} \right) 
+ {\bf N} \frac{d}{{\bar N}d\tau} - {\bf C} \frac{1}{a^2} \partial^2 + {\bf M}
\right] \,,
\end{align}
where we have defined $2\times1$ matrices
\begin{align}
& \boldsymbol{\xi} \doteq
\begin{pmatrix}
\zeta \\ \delta{Q}
\end{pmatrix} \equiv
\begin{pmatrix}
\zeta \\  \delta \chi + \frac{v}{H} \zeta 
\end{pmatrix} \,,
& \boldsymbol{\mu}^{(P)} \doteq 
\begin{pmatrix}
{ \mu}_\zeta^{(P)} \\ { \mu}_{\delta{Q}}^{(P)}
\end{pmatrix} \,,
\end{align}
and $2\times2$ matrices ${\bf N}$ and ${\bf C}$ whose nonzero components are
\begin{align}
\nonumber 
{ N}_{11} &= 
-\frac{2 \bar{V}_{\varphi }}{\dot{\bar{\varphi }}}-3 H-\frac{2\dot{H}}{H} 
- (3 w- 1) \bar{\rho} \frac{ \bar{F}_{\varphi }}{ \dot{\bar{\varphi}} \bar{F}} \,,
\\ \nonumber 
{ N}_{12} &= - \frac{1+w}{c_s^2} \frac{\bar{\rho}H}{2v} 
\left[ 
\frac{\left(1-c_s^2\right)}{H\Mp^2} 
+ \left(3 c_s^2-1\right) \frac{ \bar{F}_{\varphi }}{\dot{\bar{\varphi }}\bar{F} }
\right] 
\\ \nonumber 
&= 
- \left(\frac{1+w}{c_s^2} \frac{\bar{\rho} H^2}{\dot{\bar{\varphi }}^2 v^2} \right) {N}_{21}
\,,
\\ 
{ N}_{22} &= \frac{\dot{\bar{\varphi }}^2}{4} \left[ 
\frac{6H}{\dot{\bar{\varphi }}^2} \left(1-c_s^2\right) 
+ \left(3 c_s^2-1\right)
\frac{ \bar{F}_{\varphi }}{\dot{\bar{\varphi }}\bar{F}}
\right] \,,
\label{K-components}
\end{align}
\begin{align}\label{G-components}
&{ C}_{11} = 1 \,,
&{ C}_{22} = c_s^2 \,.
\end{align}

All components of $2\times2$ matrix ${\bf M}$ are nonzero $M_{11} \neq 0$, $M_{12} \neq 0$, $M_{21} \neq 0$, $M_{22} \neq 0$ which are shown in Appendix \ref{app-MS-marices}. The explicit values of the components of the particle source matrix $\boldsymbol{\mu}^{(P)}$, which are given by ${\mu}_\zeta^{(P)}$ and ${\mu}_{\delta{Q}}^{(P)}$, are also shown in Appendix \ref{app-MS-marices}. We will see that we do not need their explicit values to study the cosmological effects on the gravitational memory.

\section{The retarded gravitational field}\label{sec-retarded-GF}

Working with the conformal time $\eta=\int [{\bar N}(\tau)/a(\tau)] d\tau$, the background line element \eqref{BG-gravity} takes the following conformally flat form
\begin{equation}\label{BG-gravity-C}
ds^2 = a(\eta)^2 \left( - d\eta^2 + \delta_{ij} dx^i dx^j \right) \,.
\end{equation}

For the tensor perturbations, Eq. \eqref{EoM-T} gives
\begin{eqnarray}\label{EoM-hij}
\left[\partial_\eta^2 + \frac{2}{a} \frac{da}{d\eta} \partial_\eta - \partial^2 \right] h_{ij} = 4 \pi a^2 \mu_{ij} \,,
\end{eqnarray}
where we have defined normalized source $\mu_{ij} \equiv \frac{1}{2\pi\Mp^2} {\cal T}_{ij}$. 

For the scalar perturbations, Eq. \eqref{Eq-Matrix-ND} in the component form yields
\begin{align}
& \left[\partial_\eta^2 + \left( a N_{11} - \frac{1}{a} \frac{da}{d\eta} \right) \partial_\eta - \partial^2 
+ a^2 M_{11} \right] \zeta 
\nonumber \\ 
&+ \left[ a N_{12} \partial_\eta + a^2 M_{12} \right] \delta{Q} = 4\pi a^2 \mu^{(P)}_\zeta \,, \label{Eq-zeta}
\\
& \left[\partial_\eta^2 + \left( a N_{22} - \frac{1}{a} \frac{da}{d\eta} c_s^2 \right) \partial_\eta - c_s^2 \partial^2 
+ a^2 M_{22} \right] \delta{Q} 
\nonumber \\ 
&+ \left[ a N_{21} \partial_\eta + a^2 M_{21} \right] \zeta =4\pi  a^2 \mu^{(P)}_{\delta{Q}} \,. \label{Eq-Q}
\end{align}

As we have already mentioned, the memory effect for tensor perturbations $h_{ij}$ is studied in Ref. \cite{Tolish:2016ggo}. Moreover, it is shown that the field $\delta{Q}$, which corresponds to the scalar degree of freedom of the perfect fluid, does not develop any memory effect in general relativity \cite{Tolish:2016ggo}. Therefore, we only need to focus on the scalar graviton field $\zeta$. More precisely, it is expected that we only need to obtain the direct part of the retarded Green's function for $\zeta$ which is the subject of this section and Appendix \ref{app-Green}. In Appendix \ref{app-Green}, starting from the first principle and working out the fundamental solutions for the scalar modes $\zeta$ and $\delta{Q}$ in the scalar-tensor theory, we prove that $\delta{Q}$ does not develop any singularity along the lightcone of the mode $\zeta$ [Eq. \eqref{Uhat-zero-g}] while it changes the evolution of $\zeta$ inside the lightcone [Eqs. \eqref{Vhat-g}]. Therefore, $\delta{Q}$ does not contribute to the direct part of the mode $\zeta$. Then, we have systematically calculated the direct part of the corresponding retarded Green's function in Eq. \eqref{U-final}. Thus we can simply use the result Eq. \eqref{U-final} and move to the next section. 

Here in this section we obtain the same result by implementing a less rigorous but more intuitive approach which we believe is easier for the readers to follow. The readers who are only interested in the final result may simply move to the next section. On the other hand, the readers who are interested in the rigorous treatment are directed to Appendix \ref{app-Green}. 

As it is shown in Appendix \ref{app-Green}, the retarded Green's function for $\delta{Q}$ does not develop any singularities along the lightcone of $\zeta$. We thus treat the term proportional to $\delta{Q}$ in Eq. \eqref{Eq-zeta} as a source for $\zeta$ and we rewrite it in the following form
\begin{eqnarray}\label{EoM-zeta}
\left[\partial_\eta^2 + \frac{2}{A_\zeta} \frac{dA_\zeta}{d\eta} \partial_\eta - \partial^2 
+ a^2 M_{11} \right] \zeta = 4 \pi a^2 \mu_\zeta \,,
\end{eqnarray}
where we have defined
\begin{align}\label{mu}
&\mu_\zeta \equiv {\mu}^{(P)}_\zeta - \frac{1}{4\pi} \left( \frac{1}{a} N_{12} \partial_\eta + M_{12} \right) \delta{Q} \,,
\\ \label{A-zeta}
&A_\zeta(\eta) \equiv a(\eta) \exp\left[\frac{1}{2}\int_\eta 
\left( a N_{11} - \frac{3}{a} \frac{da}{d{\bar \eta}} \right)d{\bar \eta}\right] \,.
\end{align}
The corresponding Green's function then satisfies
\begin{align}
&\left[\partial_\eta^2 + \frac{2}{A_\zeta} \frac{dA_\zeta}{d\eta} \partial_\eta - \partial^2 
+ a^2 M_{11} \right] G_\zeta^{\rm ret}(x,x') 
\nonumber \\ \label{Eq-Greens}
&= \frac{4 \pi}{a^2} \delta^{(4)}(x-x') \,.
\end{align}
In general, the retarded solution for the above equation has the Hadamard representation \cite{Garabedian,Friedlander:2010eqa}
\begin{align}\label{Greens}
G_\zeta^{\rm ret}(x,x') = \left[ U_\zeta(x,x') \delta(\sigma_\zeta) + V_\zeta(x,x') \Theta(-\sigma_\zeta) \right] \Theta(t-t') \,,
\end{align}
where $U_\zeta$ and $V_\zeta$ characterize the direct and tail parts respectively and $\sigma_\zeta(x,x')$ is the geodetic interval (squared of the geodesic distance) between $x$ and $x'$ which satisfies
\begin{align}\label{sigma-zeta}
{\bar g}^{\mu\nu} {\bar \nabla}_\mu \sigma_\zeta {\bar \nabla}_\nu \sigma_\zeta 
= 2 \sigma_\zeta \,,
\end{align}
where a bar denotes that the covariant derivatives are defined in the spirit of the background metric $\bar{g}_{\mu\nu}$. Following the approach implemented in Refs. \cite{Burko:2002ge,Haas:2004kw,Poisson:2011nh}, by substituting
\begin{align}\label{Greens-change}
G_\zeta^{\rm ret}(x,x') = \frac{A_\zeta(\eta')}{A_\zeta(\eta) a(\eta')^2} g_\zeta^{\rm ret}(x,x')\,,
\end{align}
in Eq.  \eqref{Eq-Greens}, we find
\begin{align}
&\left[\partial_\eta^2 - \partial^2 + \left( a^2 M_{11} 
- \frac{1}{A_\zeta} \frac{d^2A_\zeta}{d\eta^2} \right) \right] g_\zeta^{\rm ret}(x,x')
\nonumber \\ \label{Eq-Greens-C}
&= 4 \pi \delta^{(4)}(x-x') \,.
\end{align}
The above equation is similar to the equation for the Green's function in flat spacetime. Indeed the first two terms on the left hand side of \eqref{Eq-Greens-C} correspond to the flat spacetime operator and they determine the direct part. Therefore, considering the relation between $G_\zeta^{\rm ret}$ and $g_\zeta^{\rm ret}$ in Eq. \eqref{Greens-change}, from Eq. \eqref{Greens} we find 
\begin{align}\label{U-Ubar}
U_\zeta(x,x')=\left[\frac{A_\zeta(\eta')}{A_\zeta(\eta)}\right]{\bar U}(x,x') \,,
\end{align}
where ${\bar U}$ is the corresponding quantity in the flat spacetime. Using the fact that ${\bar U}=1$, we find the direct part of the Green's function \eqref{Greens} as follows
\begin{align}\label{U}
U_\zeta(\eta,\eta') = \frac{A_\zeta(\eta')}{A_\zeta(\eta)} \,.
\end{align}
The above result coincides with the result \eqref{U-final} which is obtained from a more rigorous approach. 

Apart from the mass term, Eqs. \eqref{EoM-hij} and \eqref{EoM-zeta} have the same forms so that $A_\zeta$ plays the role of an effective scale factor for the scalar mode $\zeta$. Therefore, with the same approach, we can easily find the direct part for the tensor perturbations $h_{ij}$ as $U_h(\eta,\eta') = a(\eta')/a(\eta)$ \cite{Tolish:2016ggo}.

\section{Cosmological scalar memory effect}\label{sec-memory-effect}

In our idealized case in which GWs are produced due to the particle-like sources with energy-momentum tensor \eqref{EMT-P-J}, the memory effect is characterized by the  ``existence of first derivative of the delta functions'' in the components of the Riemann tensor \cite{Tolish:2016ggo}. The electric components of the Riemann tensor up to the linear order in tensor and scalar perturbations \eqref{Pert-gravity} are given by
\begin{align}
&\delta_1R_{i00}{}^j = \frac{1}{2} \left( \partial^2_\eta{h}_{ik} + \partial_\eta{h}_{ik} \right) \delta^{kj}
\nonumber \\ \label{Riemann-E}
& +\left( \partial_i \partial^j - \frac{1}{3} \delta_i{}^j \partial^2 \right) \Phi + \left[ \partial_\eta^2\Psi + \frac{1}{a} \frac{da}{d\eta} \left( \partial_\eta\Psi - \partial_\eta\Phi \right) \right] \delta_i{}^j \,,
\end{align}
where we have defined the Bardeen potentials \cite{Bardeen:1980kt}
\begin{align}\label{Bardeen}
&\Phi \equiv \alpha + \frac{1}{a} \partial_\eta\beta \,,
&\Psi \equiv \zeta + \frac{1}{a^2} \frac{da}{d\eta}  \beta \,.
\end{align}
For the tensor perturbations, the electric components of the Riemann tensor \eqref{Riemann-E} obviously include $\partial^2_\eta h_{ij}$ which provides first derivative of the delta function and, therefore, tensor memory effect shows up \cite{Tolish:2016ggo}. For the scalar perturbations, \eqref{Riemann-E} include $\partial^2_\eta\zeta$ while they do not include $\partial_\eta^2\delta{Q}$. The later does not provide any first derivative of the delta function and our aim is here to show that, similarly to $\partial^2_\eta h_{ij}$, the former $\partial^2_\eta\zeta$ does include first derivative of the delta function. 

Up to here, we have presented all results in the Einstein frame while to interpret the results we need to go back to the Jordan frame. In the unitary gauge $\delta\varphi=0$, which we have implemented in  \eqref{unitary-gauge}, based on \eqref{CT}, we can simply use conformal transformation ${\tilde g}_{\mu\nu} = {\bar F}^{-1} { g}_{\mu\nu}$ where only background value of conformal factor ${\bar F}$ is considered. The line element for the background \eqref{BG-gravity-C} takes the following form
\begin{align}
{\widetilde ds}^2 &= \tilde{\bar g}_{\mu\nu} dx^\mu dx^\nu 
\nonumber \\ \label{BG-gravity-C-J}
&= {\tilde a}(\eta)^2 \left( d\tau^2+ \delta_{ij} dx^i dx^j \right) , 
\hspace{.2cm} 
{\tilde a}(\eta) \equiv a(\eta) / \sqrt{{\bar F}(\eta)} \,,
\end{align}
where ${\tilde a}$ denotes the scale factor in the Jordan frame. Taking into account the change in the scale factor, we find that $\zeta$ defined by \eqref{Pert-gravity} and \eqref{unitary-gauge} does not change by the conformal transformation (see Refs. \cite{Makino:1991sg,Tsujikawa:2004my,Chiba:2008ia,Gong:2011qe,Chiba:2013mha} for the conformal invariance of the scalar perturbations in scalar-tensor theories). More precisely, the linear scalar perturbations do not change as the non-dynamical fields like $\beta$ can be simply redefined. Therefore, in the unitary gauge \eqref{unitary-gauge}, in order to go back from the Einstein frame to the Jordan frame, we only need to rewrite all results in terms of the scalar factor ${\tilde a}$ in the Jordan frame defined in \eqref{BG-gravity-C-J}. The equation of motion for $\zeta$ presented in Eq. \eqref{EoM-zeta} takes the following form in the Jordan frame
\begin{eqnarray}\label{EoM-zeta-J}
\left[\partial_\eta^2 + \frac{2}{A_\zeta} \frac{dA_\zeta}{d\eta} \partial_\eta - \partial^2 
+ {\tilde a}^2 {\tilde M}_{11} \right] \zeta = 4 \pi {\tilde a}^2 {\tilde \mu}_\zeta \,,
\end{eqnarray}
where ${\tilde M}_{11}\equiv {\bar F} {M}_{11}$, ${\tilde \mu}_{\zeta}\equiv {\bar F} {\mu}_{\zeta}$, and also
\begin{align}\label{A-zeta-J}
A_\zeta = \left(\sqrt{\bar F}{\tilde a}\right)^{3}\frac{d{\bar \varphi}}{d\eta} \exp \left[
\frac{3}{2} \int \left( \frac{w {\bar \rho}+{\bar p}_{\varphi}}{{\bar \rho}+{\bar \rho}_{\varphi}} - 1 \right) d\ln\left({\sqrt{\bar F}{\tilde a}}\right)
\right] \,,
\end{align}
in which we have used Eqs. \eqref{Hdot} and \eqref{KG-BG} when we substitute the value of $N_{11}$ defined in \eqref{K-components}. In the absence of the fluid ${\bar \rho}=0$, the mass term vanishes ${\tilde M}_{11}=0$ as it can be seen from Eq. \eqref{M-components}. We thus find the well-known result $A_\zeta \propto{\tilde a}$ for the minimally coupled ${F}=1$ massless scalar field with ${\bar p}_\varphi={\bar \rho}_\varphi$ and $d{\bar \varphi}/d\eta\propto{\tilde a}^{-2}$. There is also an apparent subtlety: for the minimal coupling and constant scalar background limit ${ F}=1$ and $\dot{\bar\varphi}=0$, as it can be seen from Eq. \eqref{mu-components}, the source does not vanish ${\tilde \mu}_\zeta = { \mu}_\zeta \neq0$. This is simply an artifact of the unitary gauge \eqref{unitary-gauge} that we have implemented. From Eq. \eqref{zeta-def-0}, we see that $\zeta = \psi$ in the unitary gauge \eqref{unitary-gauge} when $\delta\varphi=0$ while $\zeta = \frac{H}{\dot{\bar\varphi}} \delta\varphi $ if we work in the spatially flat gauge with $\psi=0$ and $E=0$. As it is shown in Appendix \ref{app-longitudinal-gauge}, there will be no source if we work with the gauge-invariant counterpart of $\delta\varphi$, which is given by $\delta\varphi + \dot{\bar \varphi} \beta$ for $E=0$, in the limit $F=1$ and $\dot{\bar\varphi}=0$. Although the $\dot{\bar\varphi}=0$ limit is not manifest in the unitary gauge, the results away from this limit are much simpler in this gauge and that is why we have implemented this gauge.

The retarded solution for $\zeta$ is given by
\begin{eqnarray}\label{zeta-sol}
\zeta(x) = \int d^4x' \sqrt{-{\bar g}(x')} G^{\rm ret}_\zeta(x,x') {\tilde \mu}_\zeta(x') \,,
\end{eqnarray}
where the retarded Green's function $G^{\rm ret}_\zeta$ is given by Eq. \eqref{Greens}. Substituting Eq. \eqref{Greens} in solution \eqref{zeta-sol} and taking into account the fact that $U_\zeta$ and $V_\zeta$ are regular functions, it can be shown that the tail part including $V_\zeta$ can at most provide singularity proportional to the delta function while the direct part including $U_\zeta$ can indeed provide singularity at the level of derivative of the delta function. Therefore, only the direct part $U_\zeta$ contributes to the memory effect. This result is completely independent of the explicit functional forms of $U_\zeta$ and $V_\zeta$. We do not repeat the corresponding analysis here, as we only need the final result in our upcoming analysis and we refer the readers to Ref. \cite{Tolish:2016ggo} for the details.

Instead of performing the integral directly in the retarded solution \eqref{zeta-sol}, the fact that the memory effect is encoded only in the direct part and the spatially flat FLRW spacetime is conformally flat, makes it possible to find a universal explicit relation between the memory effect in the FLRW spacetime and its counterpart in Minkowski spacetime. In order to do so, we first note that Eq. \eqref{U-Ubar} immediately implies
\begin{eqnarray}
\zeta^{\rm dir} (x) 
= \left[ \frac{A_\zeta(\eta_s)}{A_\zeta(\eta_o)} \right] \bar{\zeta}^{\rm dir} (x) \,,
\end{eqnarray}
where $\eta_s$ is the conformal time at the source, $\eta_o$ is the conformal time at which the detector observes the signal, and $\bar{\zeta}^{\rm dir} (x)$ is the flat spacetime counterpart of $\zeta^{\rm dir} (x)$, i.e., scalar mode characterizing the scalar field perturbations in the absence of cosmic expansion. From Eq. \eqref{Riemann-E} we have $\delta_1R_{i00}{}^j \supset \partial_\eta^2\zeta$. Taking into account the fact that the time derivatives of $A_\zeta$ do not contribute to the direct part, we find
\begin{eqnarray}\label{Riemann-memory}
\delta_1R^{\rm dir}_{({\rm S})i00}{}^j 
= \left[ \frac{A_\zeta(\eta_s)}{A_\zeta(\eta_o)} \right] \overline{\delta_1R}^{\rm dir}_{({\rm S})i00}{}^j \,,
\end{eqnarray}
where subscript ${\rm S}$ shows that only scalar perturbations are taken into account. 

The geodesic deviation equation for the deviation vector $D^\mu$, which characterizes the displacement in the detector due to the passage of the GWs, satisfies
\begin{align}\label{GDE}
& \frac{d^2D^\mu}{ds^2}
= R_{\alpha\beta\gamma}{}^\mu D^\alpha { n}^\beta { n}^\gamma \,; 
& \frac{d}{ds} \equiv n^\mu \nabla_\mu \,,
\end{align}
where $n^\mu$ is a timelike vector tangent to the geodesic trajectory which is normalized as $n_\mu n^\mu=-1$. The directional covariant derivative $d/ds$ characterizes changes along the geodesic parameterized by the affine parameter $s$. Approximating the tangent vector with its dominant background value as $n^\mu\approx {\tilde a}(\eta)^{-1} \delta^\mu{}_0$ or $d/ds \approx ({\tilde a}^{-1}) d/d\eta$ and taking into account the fact that the time derivatives of the scale factor are negligible at the time scale of interest, we find the following result at the leading order
\begin{eqnarray}
\frac{d^2D^j}{d\eta^2} = R_{i00}{}^j D^i \,.
\end{eqnarray}

Then, the displacement due to the scalar memory effect, characterized by the changes in Riemann tensor given by \eqref{Riemann-memory}, will be
\begin{eqnarray}\label{displacement}
\Delta{D}_{({\rm S})}^i =
\left[ \frac{A_\zeta(\eta_s)}{A_\zeta(\eta_o)} \right] \overline{\Delta{D}}_{({\rm S})}^i \,,
\end{eqnarray}
where $\overline{\Delta{D}}_{({\rm S})}^i$ is the displacement due to the scalar memory effect in flat spacetime, i.e., in the absence of cosmic expansion. 

Following the same steps, as it is already shown in Ref. \cite{Tolish:2016ggo}, the displacement due to the tensor memory effect is given by
\begin{eqnarray}\label{displacement-T}
\Delta{D}_{({\rm T})}^i =
\left[ \frac{a(\eta_s)}{a(\eta_o)} \right] \overline{\Delta{D}}_{({\rm T})}^i \,,
\end{eqnarray}
where $\overline{\Delta{D}}_{({\rm T})}^i$ is the displacement due to the tensor memory effect in flat spacetime. 

As it can be seen from the results \eqref{displacement} and \eqref{displacement-T}, the effects of the cosmic expansion are characterized by the values of the effective scale factor $A_\zeta$ and scale factor $a$ at the two times $\eta_s$ and $\eta_o$ for the scalar and tensor memory effects respectively. Therefore, similarly to the tensor memory effect \eqref{displacement-T}, the scalar memory effect \eqref{displacement} does not depend on the expansion history of the universe. However, depending on the coupling and mass, the value of the effective scale factor $A_\zeta$, given by Eq. \eqref{A-zeta-J}, can be different from the scale factor $a$. Thus, the detector will receive two types of scalar  \eqref{displacement} and universal tensor  \eqref{displacement-T} memory effects and, depending on the coupling and mass, the effect of the scalar memory effect can be either dominant or subleading.

The result \eqref{displacement} provides a relation between scalar memory effect in a spatially flat FLRW spacetime and its counterpart in flat spacetime for the scalar-tensor theories described by the action \eqref{action-J}. The memory effect in flat spacetime in the context of the Brans-Dicke theory is already studied in Refs. \cite{Lang:2013fna,Lang:2014osa,Hou:2020tnd,Tahura:2020vsa,Seraj:2021qja,Tahura:2021hbk}. As the action of our model Eq. \eqref{action-J} includes Brans-Dicke theory as a special case, it is straightforward to take into account the effects of cosmological expansion in the results of Refs. \cite{Lang:2013fna,Lang:2014osa,Hou:2020tnd,Tahura:2020vsa,Seraj:2021qja,Tahura:2021hbk} and also any other scalar-tensor theory which can be modeled by the action \eqref{action-J}.

\section{Summary}\label{sec-summary}

The memory effect is a permanent change in the relative separation of test particles due to the passage of GWs. In an asymptotically flat spacetime, the GWs effects can be discriminated from the gravitational tidal effects, i.e., through their different scaling near the spatial or null infinity. In the case of cosmological FLRW spacetime, which is not asymptotically flat, the situation is more subtle. In Ref. \cite{Tolish:2016ggo}, using the fact that the spatially flat FLRW spacetime is conformally flat, Tolish and Wald studied the memory effect in a universe which is filled only with a perfect fluid in general relativity. They concluded that only tensor perturbations contribute to the memory effect while scalar and vector perturbations do not. The memory effect associated to the tensor perturbations only depends on the values of the scale factor at the moment of emission from the source and at the moment of passaging the detector. In this paper, we have shown that in the context of scalar-tensor theories, the scalar perturbations associated to the scalar graviton contribute to the memory effect in the flat FLRW universe as well. The corresponding memory effect depends on the values of the ``effective scale factor for scalar graviton'' at the moment of emission from the source and at the moment of passaging the detector. The effective scale factor for the scalar graviton is given by Eq. \eqref{A-zeta-J} which reduces to the standard scale factor for the massless scalar graviton that is minimally coupled to gravity so that ${F}=\mbox{constant}$ and ${\tilde V}=\mbox{constant}$ but which is in general different from the standard scale factor. Thus, depending on the coupling and mass, the influence of the cosmic expansion on the memory effect due to the scalar perturbations can be either stronger or weaker than the one induced by the tensor perturbations.

Moreover, as a byproduct, in Appendix \ref{app-Green}, we have developed a general framework which can be used to study coupled wave equations in any curved spacetime which admits a time foliation. This will be useful not only for the studies of the cosmological memory effect but also for other scenarios which deal with solving coupled wave equations in a curved spacetime.

\begin{acknowledgments}
M.A.G. thanks the Tokyo Institute of Technology for hospitality when this work was in its final stage. The work of M.A.G. was supported by MEXT KAKENHI Grant No. 17H02890. The work of S.M. was supported in part by JSPS Grants-in-Aid for Scientific Research No.~17H02890, No.~17H06359, and by World Premier International Research Center Initiative, MEXT, Japan. The work of T.M. was supported by JST SPRING, Grant No. JPMJSP2110.
\end{acknowledgments}

\appendix

\section{Jordan frame vs. Einstein frame}\label{app-EJ-frames}

In this appendix we start with the action of a scalar-tensor theory in the Jordan frame, in which the matter is directly coupled to the metric. Performing a conformal transformation, we find the corresponding action in the Einstein frame which we use for concrete calculations in this paper. This is a quite well-known subject (see for instance Refs. \cite{Maeda:1988ab,Faraoni:1998qx,Flanagan:2004bz,Deruelle:2010ht,Chiba:2013mha} and references therein) and we present it here only to keep the paper self-contained. 

\subsection{Jordan frame}
The total action of a scalar-tensor theory in the Jordan frame, in which the matter couples directly to the metric, is as follows
\begin{align}\label{app-action-J}
&S_{\rm J}[{\tilde g},\phi,\psi] = S_{g}[{\tilde g}, \phi] + S_{\rm m}[{\tilde g},\psi] \,;
\nonumber \\
&
S_{g}[{\tilde g}, \phi]  = S_{F,{\tilde R}}[{\tilde g}, \phi] + S_{\phi}[{\tilde g},\phi] \,.
\end{align}
Here, ${\tilde g}_{\mu\nu}$ and $\phi$ are the metric in the Jordan frame and a scalar field which are the dynamical variables in the gravity sector, while $\psi$ collectively represents all fields and particles which are present in the system under consideration. The gravitational part of the action is defined by
\begin{align}\label{app-action-g}
&S_{F,{\tilde R}}[{\tilde g}, \phi] = \frac{\Mp^2}{2} \int d^4x\sqrt{-{\tilde g}} F(\phi) {\tilde R}\,,
\nonumber \\
&
S_{\phi}[{\tilde g}, \phi] = - \int d^4x\sqrt{-{\tilde g}} 
\left[
\frac{1}{2} {\tilde K}(\phi) {\tilde g}^{\mu\nu}\partial_{\mu}\phi \partial_{\nu}\phi + {\tilde V}(\phi)
\right] ,
\end{align}
where $\Mp=(8\pi{G})^{-1/2}$ is the reduced Planck mass, ${\tilde R}$ is the Ricci scalar in the Jordan frame, $F$, ${\tilde K}$, ${\tilde V}$ are functions of the scalar field. The matter action is given by
\begin{eqnarray}\label{app-action-m}
S_{\rm m}[{\tilde g},\psi] = \int d^4x\sqrt{-{\tilde g}} {\mathcal L}_{\rm m}({\tilde g}_{\alpha\beta},\psi) \,,
\end{eqnarray}
in which ${\mathcal L}_{\rm m}$ is the Lagrangian density of the matter sector.

Taking variation of the action \eqref{app-action-J} with respect to the metric and using the relation
\begin{align}
\delta_{\tilde g} S_{F,{\tilde R}}[{\tilde g}, \phi] &= \frac{\Mp^2}{2} \int d^4x \sqrt{-{\tilde g}} 
\bigg[ 
F(\phi) {\tilde G}_{\mu\nu} 
\nonumber \\ \nonumber
&
+ {\tilde g}_{\mu\nu} {\tilde \Box}F 
- {\tilde \nabla}_\mu {\tilde \nabla}_\nu F 
\bigg] 
\delta{\tilde g}^{\mu\nu} \,,
\end{align}
we find the Einstein equations
\begin{eqnarray}\nonumber
\Mp^2 F(\phi) {\tilde G}_{\mu\nu} = {\tilde T}^{(\phi)}_{\mu\nu} 
+ \Mp^2  {\tilde \nabla}_\mu {\tilde \nabla}_\nu F - \Mp^2  {\tilde g}_{\mu\nu} {\tilde \Box}F
+ {\tilde T}^{({\rm m})}_{\mu\nu} \,,
\end{eqnarray}
where
\begin{align}
{\tilde T}^{(\phi)}_{\mu\nu} 
&= \frac{-2}{\sqrt{-\tilde g}} \frac{\delta{S}_{\phi}}{\delta{\tilde g}^{\mu\nu}}
\nonumber \\
&= {\tilde K}(\phi) \partial_\mu\phi \partial_\nu\phi 
- {\tilde g}_{\mu\nu} 
\left[
\frac{1}{2} {\tilde K}(\phi) {\tilde g}^{\alpha\beta} \partial_\alpha\phi \partial_\beta\phi + {\tilde V}(\phi) 
\right] \,,
\\ \label{EMT-J}
{\tilde T}^{({\rm m})}_{\mu\nu} 
&= \frac{-2}{\sqrt{-\tilde g}} \frac{\delta{S}_{\rm m}}{\delta{\tilde g}^{\mu\nu}} \,.
\end{align}
Taking variation of the action \eqref{app-action-J} with respect to $\phi$, we find
\begin{align}
&\delta_{\phi} S_{\rm J}[{\tilde g}, \phi,\psi] = \delta_{\phi} S_{g}[{\tilde g}, \phi]
\nonumber \\
&= \left[ {\tilde K}(\phi) {\tilde \Box}\phi
+ \frac{{\tilde K}_{,\phi}}{2} {\tilde g}^{\alpha\beta}  \partial_\alpha\phi \partial_\beta\phi 
- {\tilde V}_{,\phi} + \frac{\Mp^2}{2} F_{,\phi} {\tilde R}  \right] \delta\phi = 0 \,,
\end{align}
where we have used the fact that the matter action is independent of the scalar field in the Jordan frame. We also have the conservation of the matter field
\begin{eqnarray}
\tilde{\nabla}_\mu {\tilde T}^{({\rm m})\,\mu}{}_{\nu} = 0 \,.
\end{eqnarray}

\subsection{Einstein frame}

In order to go to the Einstein frame, we perform the following conformal transformation
\begin{equation}\label{app-CT}
{\tilde g}_{\mu\nu} = F(\phi)^{-1} g_{\mu\nu} \,,
\end{equation}
where $g_{\mu\nu}$ is the metric in the Einstein frame. We thus have
\begin{eqnarray}\label{app-CT-2}
&&{\tilde g}^{\mu\nu} = F\, g^{\mu\nu} \,, \hspace{1cm}
\sqrt{-{\tilde g}} = F^{-2} \, \sqrt{-g} \,, \nonumber \\ \nonumber
&& {\tilde R} = F \left[
R + 3 \frac{F_{,\phi}}{F} {\tilde g}^{\mu\nu} \nabla_{\mu}\nabla_{\nu} \phi 
+ 3 {\tilde g}^{\mu\nu}\partial_{\mu}\phi \partial_{\nu}\phi 
\left(  \frac{F_{,\phi\phi}}{F} - \frac{3}{2} \frac{F_{,\phi}^2}{F^2} \right)
\right] \,,
\end{eqnarray}
where $R$ is the Ricci scalar associated to $g_{\mu\nu}$. 

Then, the gravitational action in the Einstein frame takes the form
\begin{eqnarray}\label{app-action-E0}\nonumber
S_{g}[{g}, \phi] = \int d^4x\sqrt{-{ g}} 
\left[
\frac{\Mp^2}{2} { R} 
- \frac{1}{2} { K}(\phi) { g}^{\mu\nu}\partial_{\mu}\phi \partial_{\nu}\phi - { V}(\phi)
\right] \,,
\end{eqnarray}
where we have defined
\begin{eqnarray}
K \equiv \frac{{\tilde K}}{F} + \frac{3}{2} \Mp^2 \left( \frac{F_{,\phi}}{F} \right)^2 \,, \hspace{1cm}
V \equiv \frac{{\tilde V}}{F^2} \,.
\end{eqnarray}

Redefining the scalar field as 
\begin{equation}\label{varphi}
\varphi = \int \sqrt{K(\phi)} \, d\phi \,,
\end{equation}
the kinetic term of the scalar field takes the canonical form and we find the following total action in the Einstein frame
\begin{align}
S_{\rm E}[{g}, \varphi,\psi] &= \int d^4x\sqrt{-{ g}} 
\bigg[
\frac{\Mp^2}{2} { R} 
\nonumber \\ \label{app-action-E}
&- \frac{1}{2} { g}^{\mu\nu}\partial_{\mu}\varphi \partial_{\nu}\varphi - { V}(\varphi)
\bigg]
+ S_{\rm m}[\tilde{g}, \psi]  \,,
\end{align}
where it is understood that $\tilde{g}$ in \eqref{app-action-E} is given by \eqref{app-CT} and that $\phi$ is considered as a function of $\varphi$.

In the absence of matter, actions \eqref{app-action-J} and \eqref{app-action-E} are completely equivalent at the classical level. As we see, the matter action depends on the field in the Einstein frame while it was independent of the scalar field in the Jordan frame. That is why the scalar-tensor theories are treated as modified gravity theories and they are not simply general relativity plus a minimally coupled scalar field. 

Varying the action \eqref{app-action-E} with respect to $g^{\mu\nu}$, we find the Einstein equations in the Einstein frame
\begin{equation}\label{app-EE}
\Mp^2 G_{\mu\nu} = T^{(\varphi)}_{\mu\nu} + T^{({\rm m})}_{\mu\nu} \,,
\end{equation}
where 
\begin{eqnarray}\label{app-EMT-phi}
T^{(\varphi)}_{\mu\nu} = \partial_{\mu} \varphi \partial_{\nu} \varphi
- g_{\mu\nu} 
\left[
\frac{1}{2} g^{\mu\nu}\partial_{\mu}\varphi \partial_{\nu}\varphi + V(\varphi) 
\right] \,,
\end{eqnarray}
\begin{eqnarray}\label{app-EMT-m}
T^{({\rm m})}_{\mu\nu} = \frac{-2}{\sqrt{- g}} \frac{\delta{S}_{\rm m}}{\delta{ g}^{\mu\nu}} \,.
\end{eqnarray}

We need to take into account the dependence of $S_{\rm m}$ on the scalar field when we take the variation with respect to the scalar field. We, however, note that the dependence on the scalar field only comes through the conformal transformation \eqref{app-CT} and we can simply use the chain rule to take the variation. Then, taking variation of the action \eqref{app-action-E} with respect to the scalar field $\varphi$ we find the equation of motion for the scalar field as follows
\begin{equation}\label{app-EOM-varphi}
\Box{\varphi} - V_{,\varphi} = \frac{F_{,\varphi}}{2F} \, T^{({\rm m})} \,,
\end{equation}
where $T^{({\rm m})} = g^{\alpha\beta} T^{({\rm m})}_{\alpha\beta}$ is the trace of the matter energy-momentum tensor in the Einstein frame. 

Moreover, from the Bianchi identity we know $\nabla_{\mu} G^{\mu}{}_{\nu} = 0$ which implies $\nabla_{\mu} \big( T^{(\varphi)\mu}{}_{\nu} + T^{({\rm m})\mu}{}_{\nu} \big) = 0$. Using the explicit form of the scalar field energy-momentum tensor \eqref{app-EMT-phi}, we find $\nabla_{\mu} T^{(\varphi)\mu}{}_{\nu} = - \big( \Box{\varphi} - V_{,\varphi} \big) \nabla_{\nu}\varphi$ which after using Eq. \eqref{app-EOM-varphi} leads to the following conservation equation for the matter
\begin{eqnarray}\label{CoE}
\nabla_{\mu} T^{({\rm m})\mu}{}_{\nu} = - \frac{F_{,\varphi}}{2F} \, T^{({\rm m})}  \nabla_{\nu}\varphi \,.
\end{eqnarray}

Note that we did not assume any form for the matter and all results found in this appendix can be applied to any type of matter. Note also that if there are different contributions to the matter energy-momentum tensor, energy-momentum of each type of matter are separately conserved.

\section{Matter energy-momentum tensor in different frames}\label{app-EJ-EMT}

We have considered a general matter field in the previous appendix. Here we study in detail two types of matter sources with which we deal in our setup: perfect fluids and relativistic particles. We find relations between physical quantities in the Jordan and Einstein frames for these two types of matter sources.

Using Eq. \eqref{app-CT}, we can rewrite the energy-momentum tensor in the Einstein frame $T^{({\rm m})}_{\mu\nu}$, defined in Eq. \eqref{app-EMT-m}, as follows
\begin{align}\label{app-EMT-EJ}
&T^{({\rm m})}_{\mu\nu} \big( g, \varphi, \psi \big)
= \frac{1}{F(\varphi)} \, {\tilde T}^{({\rm m})}_{\mu\nu} \big( {\tilde g}, \psi \big) \,;
\\ \nonumber 
&{\tilde T}^{({\rm m})}_{\mu\nu} = 
-\frac{2}{\sqrt{-{\tilde g}}} \frac{\delta (\sqrt{-{\tilde g} } {\mathcal L}_{\rm m}) }{\delta {\tilde g}^{\mu\nu}} \,,
\end{align}
where ${\tilde T}^{({\rm m})}_{\mu\nu}$ is the energy-momentum tensor in the Jordan frame defined in Eq. \eqref{EMT-J}. This relation is true for any type of matter.

In the following subsections, we separately study the special cases of perfect fluid and particle-like sources.

\subsection{Perfect fluid}

In the Jordan frame, the perfect fluid energy-momentum tensor ${\tilde T}^{({\rm F})}_{\mu\nu}$ is given by
\begin{align}\label{app-EMT-F-J}
&{\tilde T}^{({\rm F})}_{\mu\nu}\big({\tilde g},{\tilde U}\big) = 
({\tilde \rho}+{\tilde p}) {\tilde U}_\mu {\tilde U}_\nu + {\tilde p}\, {\tilde g}_{\mu\nu} \,;
\\ \nonumber 
&
{\tilde g}^{\mu\nu}{\tilde U}_\mu{\tilde U}_\nu = -1 \,,
\end{align}
where ${\tilde U}_\mu$, ${\tilde \rho}$, and ${\tilde p}$ are four-velocity, energy density, and pressure in the Jordan frame respectively. 

In the Einstein frame, we also have 
\begin{align}\label{app-EMT-F-E}
&{ T}^{({\rm F})}_{\mu\nu}\big({ g},{ U}\big) = 
({ \rho}+{ p}) { U}_\mu { U}_\nu + { p}\, { g}_{\mu\nu} \,;
\\ \nonumber 
&{ g}^{\mu\nu}{ U}_\mu{ U}_\nu = -1 \,,
\end{align}
where ${ U}_\mu$, ${ \rho}$, and ${ p}$ are four-velocity, energy density, and pressure in the Einstein frame respectively. 

In order to find the relation between the quantities in two frames, we first note that the normalization of the four-velocities in Eqs. \eqref{app-EMT-F-J} and \eqref{app-EMT-F-E} implies
\begin{eqnarray}\label{app-U-EJ}
U_\mu \equiv \sqrt{F} \, {\tilde U}_\mu \,. 
\end{eqnarray}
Using the above result and the relation \eqref{app-EMT-EJ}, we find
\begin{eqnarray}\label{app-rho-p-EJ}
\rho = F^{-2}\, {\tilde \rho} \,, \hspace{1cm} p = F^{-2}\, {\tilde p} \,.
\end{eqnarray}
The above results for the perfect fluid are already well known and we just presented them for the sake of completeness.

\subsection{Massive and massless particles}

For the particle energy-momentum tensor, as far as we know, the relations between the physical quantities in the Jordan and Einstein frames are not completely derived yet. We, thus, present them here in some details. We start with the action of relativistic particles which is invariant under a conformal transformation. We, therefore, can systematically find relations between physical quantities before and after the conformal transformation.

In the Jordan frame, the action for a relativistic particle is given by
\begin{eqnarray}\label{action-particle-J}
S_{{\rm J}}^{(P)} = \frac{1}{2} \int d{\tilde \lambda} 
\left[
\frac{1}{{\tilde e}} 
\left( {\tilde g}_{\alpha\beta} \frac{dx^\alpha}{d{\tilde \lambda}} \frac{dx^\beta}{d{\tilde \lambda}} 
\right)
- {\tilde e} \, {\tilde m}^2 
\right] \,,
\end{eqnarray}
where ${\tilde m}$ is the mass of the particle, ${\tilde \lambda}$ is an arbitrary curve parameter, and ${\tilde e}$ is an auxiliary field. 

Let us first focus on the case of a massive particle with ${\tilde m}\neq0$. In this case, the solution for the auxiliary field ${\tilde e}$ can be obtained from its equation of motion as follows
\begin{eqnarray}\label{e-tilde-sol}
{\tilde e} = \frac{1}{{\tilde m}} \sqrt{-{\tilde g}_{\alpha\beta} \frac{dx^\alpha}{d{\tilde \lambda}} \frac{dx^\beta}{d{\tilde \lambda}}} \,.
\end{eqnarray}
Substituting the above result in the action \eqref{action-particle-J}, we find the following action for the massive particles
\begin{eqnarray}\label{action-particle-J-massive}
S_{{\rm J}}^{(M)} = - {\tilde m} \int d{\tilde \lambda} \,
\sqrt{ - {\tilde g}_{\alpha\beta} \frac{dx^\alpha}{d{\tilde \lambda}} \frac{dx^\beta}{d{\tilde \lambda}} } \,.
\end{eqnarray}
For the massive particles the curve parameter can be simply chosen as the proper time ${\tilde \lambda} = {\tilde \tau}$.

In the case of a massless (null) particle with ${\tilde m}=0$, the action \eqref{action-particle-J} reduces to
\begin{eqnarray}\label{action-particle-J-null}
S_{{\rm J}}^{(N)} = \frac{1}{2} \int d{\tilde \lambda}
\frac{1}{{\tilde e}} 
\left( {\tilde g}_{\alpha\beta} \frac{dx^\alpha}{d{\tilde \lambda}} \frac{dx^\beta}{d{\tilde \lambda}} 
\right) \,.
\end{eqnarray}
Contrary to the case of the massive particle, the auxiliary field ${\tilde e}$ cannot be fixed by its equation of motion. Instead, the equation of motion of ${\tilde e}$ gives the Hamiltonian constraint 
\begin{eqnarray}
{\tilde g}_{\alpha\beta} \frac{dx^\alpha}{d{\tilde \lambda}} \frac{dx^\beta}{d{\tilde \lambda}} = 0 \,.
\end{eqnarray}
The curve parameter ${\tilde \lambda}$ then can be chosen as an affine parameter. However, one can keep ${\tilde e}$ and work with ${\tilde e}d{\tilde \lambda}$ as an affine parameter. In this case, freedom in ${\tilde e}$ allows us to keep the setup invariant under reparameterization of ${\tilde e}d{\tilde \lambda}$.

In the Einstein frame, the action takes the form
\begin{eqnarray}\label{action-particle-E}
S_{{\rm E}}^{(P)} = \frac{1}{2} \int d{ \lambda} 
\left[
\frac{1}{{ e}} 
\left( { g}_{\alpha\beta} \frac{dx^\alpha}{d{ \lambda}} \frac{dx^\beta}{d{ \lambda}} 
\right)
- { e} \, { m}^2 
\right] \,,
\end{eqnarray}
where $m$ denotes the mass of the relativistic particle in the Einstein frame. 

Looking at the first terms in the right hand sides of the actions \eqref{action-particle-J} and \eqref{action-particle-E}, from conformal transformation \eqref{app-CT} we find the following relation
\begin{eqnarray}\label{edlambda-JE}
{\tilde e}\, d{\tilde\lambda} = F^{-1}\, e\, d\lambda \,.
\end{eqnarray}
Using the above result and comparing the second terms on the right hand sides of the actions \eqref{action-particle-J} and \eqref{action-particle-E}, we find the following relation between the mass of the particles in the Jordan and Einstein frames
\begin{eqnarray}\label{m-JE}
{\tilde m} = \sqrt{F}\, m \,.
\end{eqnarray}

In the case of the massive particle, when ${\tilde e}$ is fixed by its equation of motion \eqref{e-tilde-sol}, 
we choose the curve parameter to be the proper time of the massive particle as ${\tilde \lambda} = {\tilde \tau}$. Then, from Eqs. \eqref{e-tilde-sol} and \eqref{edlambda-JE} we find the following results
\begin{eqnarray}\label{tau-JE}
{\tilde e} = \frac{e}{\sqrt{F}} , \hspace{.2cm} d{\tilde \tau} = \frac{d\tau}{\sqrt{F}} ;
\hspace{.2cm} 
{ e} = \frac{1}{{ m}} \sqrt{-{ g}_{\alpha\beta} \frac{dx^\alpha}{d{ \lambda}} \frac{dx^\beta}{d{ \lambda}}} .
\end{eqnarray}

Finally, taking variation of the actions \eqref{action-particle-J-massive} and \eqref{action-particle-J-null} with respect to the metric ${\tilde g}_{\mu\nu}$ we find
\begin{eqnarray}\label{app-EMT-M-J}
{\tilde T}^{(M)}_{\mu\nu} &=& 
{\tilde m} {\tilde u}_\mu {\tilde u}_\nu
\frac{1}{\sqrt{-{\tilde g}}} \frac{d{\tilde \tau}}{dt} \,
\delta^{(3)}({\bf x}-{\bf z}(t)) \,,
\\ \label{app-EMT-N-J}
{\tilde T}^{(N)}_{\mu\nu} &=& 
{\tilde k}_\mu {\tilde k}_\nu 
\frac{1}{\sqrt{-{\tilde g}}} \frac{{\tilde e}d{\tilde \lambda}}{dt} \,
\delta^{(3)}({\bf x}-{\bf y}(t)) \,,
\end{eqnarray}
where ${\tilde u}^\mu=\frac{d x^{\mu}}{d{\tilde \tau}}$ is the four-velocity of the massive particles and ${\tilde k}^{\mu}=\frac{d x^{\mu}}{{\tilde e}d{\tilde \lambda}}$ is the four-momentum of the massless particles in the Jordan frame. In the Einstein frame, we have 
\begin{eqnarray}\label{app-EMT-M-E}
{ T}^{(M)}_{\mu\nu} &=& 
{ m} { u}_\mu { u}_\nu
\frac{1}{\sqrt{-{ g}}} \frac{d{ \tau}}{dt} \,
\delta^{(3)}({\bf x}-{\bf z}(t)) \,,
\\ \label{app-EMT-N-E}
T^{(N)}_{\mu\nu} &=& 
{ k}_\mu { k}_\nu 
\frac{1}{\sqrt{-{ g}}} \frac{{ e}d{ \lambda}}{dt} \,
\delta^{(3)}({\bf x}-{\bf y}(t)) \,,
\end{eqnarray}
where ${ u}^\mu= \frac{d x^{\mu}}{d{ \tau}}$ is the four-velocity of the massive particles and ${ k}^\mu = \frac{d x^{\mu}}{{ e}d{ \lambda}}$ is the four-momentum of the massless particle in the Einstein frame. 

For the massive particle, using the results \eqref{tau-JE}, we find
\begin{eqnarray}\label{u-JE}
{\tilde u}^\mu = \sqrt{F} { u}^\mu \,, 
\hspace{1cm} 
{\tilde u}_\mu = \frac{{ u}_\mu}{\sqrt{F}} \,,
\end{eqnarray}
while for massless particle we find
\begin{eqnarray}\label{k-JE}
{\tilde k}^\mu = F \, { k}^\mu \,, 
\hspace{1cm} 
{\tilde k}_\mu = { k}_\mu \,.
\end{eqnarray}
Here, we started from the first principle and we worked with the action. However, looking at the geodesic equation, it is also possible to find relation between affine parameters ${\tilde e}d{\tilde \lambda}$ and ${ e}d{ \lambda}$ at the level of equation of motion \cite{Wald:1984rg}. Indeed, it is straightforward to show that if a massless particle with four-momentum ${\tilde k}_{\mu}$ satisfies the geodesic equation ${\tilde k}^\alpha {\tilde \nabla}_{\alpha} {\tilde k}_{\mu} = 0$ with affine parameter ${\tilde e}d{\tilde \lambda}$ in the Jordan frame, it also satisfies geodesic equation ${ k}^\alpha { \nabla}_{\alpha} { k}_{\mu} = 0$ with the affine parameter $e d\lambda$ in the Einstein frame.

\section{Explicit values of mass matrix and source vector}\label{app-MS-marices}

We present explicit forms of the components of the mass and source matrices defined in Eq. \eqref{Eq-Matrix-ND}. Indeed, we do not need these explicit results for our purpose in this paper and we only present them here for the sake of completeness.

\begin{widetext}
The components of the mass matrix ${\bf M}$ are given by
\begin{align}
M_{11} & = \frac{(w+1) \bar{\rho} }{4\Mp^2 c_s^2}
\Bigg\{
2 \epsilon \left( 3 c_s^2 - 1 \right)
- 6 c_s^2 \left( 1 + c_s^2 \right)
+\frac{(1+w)\bar{\rho } \left(1-c_s^2\right)}{H^2 \Mp^2}
-\frac{4c_s^2\bar{V}_{\varphi }}{H \dot{\bar{\varphi}}}
+ \frac{\left(3 c_s^2-1\right){}^2\Mp^2\bar{F}_{\varphi }^2}{\bar{F}^2}
\nonumber \\
& + \left[
\dot{\bar{\varphi }}^2 \left(3c_s^4-4 c_s^2+1\right)-\bar{\rho } \left((3 w-5)c_s^2+w+1\right)-2H^2\Mp^2 \left(3 c_s^2-1\right) \left(\epsilon -3c_s^2\right)
\right] \frac{\bar{F}_{\varphi }}{H \dot{\bar{\varphi }} \bar{F}}
\Bigg\}
\,, \nonumber \\
M_{12} &= \frac{(w+1) \bar{\rho }}{4 \Mp^2} 
\Bigg\{
3\left(1-3 c_s^2\right) \frac{\dot{\bar{\varphi }}
	\bar{F}_{\varphi }}{2 H \bar{F}}-\frac{\dot{\bar{\varphi
	}}^2 \left(c_s^2-1\right)}{H^2 \Mp^2}+15 c_s^2-4 \epsilon +3
\Bigg\} 
\,, \nonumber \\
M_{21} &= \frac{1}{2v} \Bigg\{
\epsilon H \left(3+4\epsilon-9 c_s^2\right)
- \frac{(w+1) \bar{\rho } \dot{\bar{\varphi }}^2\left(c_s^2-1\right)}{2 H^3 \Mp^4}
-\frac{(w+1) \bar{\rho }\left(-9 c_s^2+4 \epsilon +3\right)+12 \dot{\bar{\varphi}}^2}{2 H \Mp^2}
\nonumber \\
&
+ \left[\left(1-3 c_s^2\right)\frac{\bar{F}_{\varphi }}{H \bar{F}}-\frac{2\dot{\bar{\varphi}}}{H^2 \Mp^2}\right] \bar{V}_{\varphi } 
- \left(3 c_s^2-1\right) \left(7 \dot{\bar{\varphi }}^2+2(3 w-1) \bar{\rho }\right) \frac{ \bar{F}_{\varphi }^2}{2H \bar{F}^2}
\nonumber \\
&
+ \left[(5\epsilon -3) \left(3 c_s^2-1\right)-\frac{\bar{\rho } \left(3(w+1) c_s^2+11 w-5\right)}{2 H^2 \Mp^2}\right] \frac{\dot{\bar{\varphi }} \bar{F}_{\varphi } }{2\bar{F}} 
+ \left(3c_s^2-1\right) \frac{\dot{\bar{\varphi }}^2 \bar{F}_{\varphi \varphi }}{H\bar{F}}
\Bigg\}  \,, \nonumber \\
M_{22} &= \frac{(w+1) \bar{\rho }}{4 \Mp^2}
\Bigg[
3-4 \epsilon +15 c_s^2
+ \left(1-c_s^2\right) \frac{\dot{\bar{\varphi}}^2}{H^2 \Mp^2}
+ 3\left(1-3 c_s^2\right) \frac{\dot{\bar{\varphi }} \bar{F}_{\varphi }}{2 H \bar{F}}
\Bigg]
\,, \label{M-components}
\end{align}
where we have defined $\epsilon\equiv - \dot{H}/H^2$.

The components of the particle source matrix $\boldsymbol{\mu}^{(P)}$ are also given by
\begin{align}
\nonumber 
{\mu}_\zeta^{(P)} 
&= - \frac{1}{4\pi\Mp^2} \left\{
\alpha^{(P)} + \psi^{(P)} + \frac{2}{3} \partial^2 E^{(P)}
-
\left[ (3 - \epsilon) H + \frac{\bar{V}_{\varphi}}{\dot{\bar{\varphi }}} - \frac{(1+w)\bar{\rho}}{4 H \Mp^2 c_s^2} \left(1-c_s^2\right)
\right]
\beta^{(P)} \right\} 
\\ \nonumber 
& + \frac{H \bar{F}_{\varphi }}{4\pi\dot{\bar{\varphi}} \bar{F} } 
\left[ \alpha^{(P)} + 3 \psi^{(P)}
+ \frac{ (1+w) \bar{\rho } }{4 H \Mp^2 c_s^2} 
\left(1 - \frac{5-3 w}{1+w} c_s^2\right) \beta ^{(P)} \right] \,,
\\ \nonumber 
\mu_{\delta{Q}}^{(P)} 
&= \frac{v}{4\pi {H} \Mp^2}
\Bigg\{
c_s^2 \alpha^{(P)}+ \psi^{(P)} + \frac{2}{3} \partial^2 E^{(P)}
\\
&
-
\frac{1}{4} \bigg[
\left( 3 - 4 \epsilon + 15 c_s^2 \right) H
+ \dot{\bar{\varphi }}^2 \left(
\frac{ \left(1-c_s^2\right)}{H\Mp^2}
- \frac{3 }{2} \left(3 c_s^2-1\right) \frac{ \bar{F}_{\varphi }}{\dot{\bar{\varphi }} \bar{F}} \right) 
\bigg]
\beta^{(P)}
\Bigg\} \,.  \label{mu-components}
\end{align}

\end{widetext}


\section{The minimal coupling and constant scalar background limit}\label{app-longitudinal-gauge}

From Eq. \eqref{EOM-varphi}, we see that for the case of minimally coupled scalar field with $F=\mbox{constant}$ and the constant scalar field background $\dot{\bar\varphi}=0$, there should be no source for the scalar field $\varphi$ and, therefore, there should be no scalar memory effect in this case. On the other hand, if we take the limit $F=\mbox{constant}$ and $\dot{\bar\varphi}=0$ in Eq. \eqref{mu-components}, we find ${\mu}_\zeta^{(P)} \neq0$ for the mode $\zeta$. One may concern that there would be a scalar memory effect in this limit which is apparently not consistent with Eq. \eqref{EOM-varphi} that holds at the fully non-linear level. In this appendix, we clarify that this is simply a gauge artifact and, using a gauge-invariant counterpart of the scalar field perturbation, we show that the limit is indeed consistent. The reason why we used the unitary gauge \eqref{unitary-gauge} in the paper is that the results away from the constant scalar background become very simple in this gauge.

For the gravity and fluid sectors, we deal with scalar perturbations $\{ \alpha,\beta,\psi,E,\delta\phi \}$ and $\{\delta\rho,U\}$ respectively. We have worked with variable $\zeta$ defined in Eq. \eqref{zeta-def-0}
\begin{eqnarray}
\zeta = \psi - \frac{H}{\dot{\bar\varphi}} \delta\varphi \,.
\end{eqnarray}
The quantity $\zeta$ is a combination of $\psi$ and $\delta\varphi$ which becomes $\zeta=\psi$ in unitary gauge \eqref{unitary-gauge} when $\delta\varphi=0$ while $\zeta = \frac{H}{\dot{\bar\varphi}} \delta\varphi $ if we fix the gauge as $\psi=0$. In order to make the minimal coupling and constant scalar background limit $F=\mbox{constant}$ and $\dot{\bar\varphi}=0$ manifest in the equations, we need to work with $\delta\varphi$. In order to avoid any gauge artifact, let us look at the following gauge-invariant variables\footnote{Remember that all perturbations in the particle energy-momentum tensor are gauge-invariant.}
\begin{align}\label{gauge-invariant-Psi-Phi}
&\Phi = \alpha + ( \beta - a^2 \dot{E} )\dot{} \,,
&\Psi = \psi + H ( \beta - a^2 \dot{E} ) \,,
\\ \label{gauge-invariant-dphi}
&\delta\varphi_{\rm N} = \delta\varphi + \dot{\bar \varphi}
( \beta - a^2 \dot{E} ) \,,
\\
&\delta\rho_{\rm N} = \delta\rho + \dot{\bar \rho} 
( \beta - a^2 \dot{E} ) \,,
&U_{\rm N} = U + ( \beta - a^2 \dot{E} ) \,.
\end{align}
In terms of the gauge-invariant variables, $\zeta$ becomes
\begin{eqnarray}\label{zeta-def}
\zeta = \Psi - \frac{H}{\dot{\bar\varphi}} \delta\varphi_{\rm N} \,,
\end{eqnarray}
which shows that, independent of the gauge, $\zeta$ always includes metric perturbations. This clarifies the apparent discrepancy that we mentioned in the beginning of this appendix. Therefore, we need to work with $\delta\varphi_{\rm N}$ to make the minimal coupling and constant scalar background limit $F=\mbox{constant}$ and $\dot{\bar\varphi}=0$ manifest in the equations. 

After defining gauge-invariant variables, in any gauge we will deal with $\delta\varphi_{\rm N}$ in terms of which the limit $F=\mbox{constant}$ and $\dot{\bar\varphi}=0$ will be manifest. So, let us fix the gauge for the sake of simplicity and work in the {\it longitudinal gauge}
\begin{equation}
\beta = 0 \,, \hspace{1.5cm} E = 0 \,.
\end{equation}
We thus find the following gauge-fixed quantities
\begin{eqnarray}
&	\Phi = \alpha \,,
	\hspace{.5cm}
	\Psi = \psi \,,
	\hspace{.5cm}
	\delta\varphi_{\rm N} = \delta\varphi \,,
\\
& 
	\delta\rho_{\rm N} = \delta\rho \,,
	\hspace{.5cm}
	U_{\rm N} = U \,.
\end{eqnarray}
As it is clear, in this gauge, $\delta\varphi$, $\delta\rho$, $U$ are gauge-invariant and, from now on, we will drop the subscript and show $\delta\varphi_{\rm N}$, $\delta\rho_{\rm N}$, $U_{\rm N}$ by $\delta\varphi, \delta\rho, U$.

The line element for the scalar perturbations in this gauge takes the form
\begin{equation}
	ds^2 = - \bar{N}^2 ( 1 + 2 \Phi ) dt^2 + a^2 (1+2\Psi) \delta_{ij} dx^i dx^j \,.
\end{equation}

The linearized Einstein Eqs. \eqref{EE-ST} give
\begin{align}\nonumber
	&2 \Mp^2 \left[ 3 H ( \dot{\Psi } - H \Phi ) - \frac{1}{a^2} \partial^2 \Psi \right]
\\ \label{EEp-00}
	&= - \dot{\bar{\varphi }}^2 \Phi 
	+ \dot{\bar{\varphi }} \dot{\delta \varphi } 
	+ \bar{V}_{\varphi} \delta \varphi  + \delta \rho + 2 \alpha^{(P)} \,,
\end{align}
\begin{align}\label{EEp-0i}
	- 2 \Mp^2 ( \dot{\Psi} - H \Phi ) 
	= \dot{\bar{\varphi }} \delta \varphi + (1+w) \bar{\rho }\, { U} + \beta^{(P)} \,,
\end{align}
\begin{align}\label{EEp-ij}
	\frac{1}{a^2} \Mp^2 \partial^i \partial_j (\Phi + \Psi)= -2 \partial^i \partial_j E^{(P)} \,, \hspace{1cm} i \neq j
\end{align}
\begin{align}\label{EEp-ii}
	&2 \Mp^2 \left[
	\ddot{\Psi} + H ( 3 \dot{\Psi } - \dot{\Phi } )
	- ( 2 \dot{H} + 3 H^2 ) \Phi 
	\right]
	\nonumber \\
	&- \frac{2}{3} \Mp^2 \frac{1}{a^2} \partial^2 (\Phi + \Psi) 
	\nonumber \\
	& = \dot{\bar{\varphi }}^2 \Phi - \dot{\bar{\varphi }} \dot{\delta \varphi }
	+ \bar{V}_{\varphi } \delta \varphi
	- c_s^2 \delta \rho + 2 \psi^{(P)} \,, \hspace{1cm} i = j
\end{align}

Linearizing equations for the conservation of perfect fluid \eqref{EOM-COM-F}, we find
\begin{align}\label{COMF-p-t}
&\frac{\dot{\delta\rho}}{\bar{\rho }} 
	+ 3H \left( 1 +	c_s^2 \right) \frac{\delta\rho}{\bar{\rho }} 
	+ (1+w)\Big( 3 \dot{\Psi } - \frac{1}{a^2} \partial^2 U \Big)
\nonumber 
\\
	&= - \frac{\dot{\bar{\varphi }}\bar{F}_{\varphi } }{2 \bar{F}} (1 - 3 c_s^2) \frac{\delta\rho}{\bar{\rho }}
- \frac{1}{2} (1-3 w) 
\frac{d}{{\bar N}dt}
\left( \frac{\bar{F}_{\varphi}}{\bar{F}} {\delta \varphi }\right) \,,
\end{align}
\begin{align}\label{COMF-p-i}
&\dot{ U} + 3H { U}  - \left( \frac{\dot{\bar \rho}}{\bar{\rho}} + \frac{\dot{w}}{1+w} \right) { U}
\\ \nonumber 
& 
= \frac{c_s^2}{(1+w)} \frac{\delta\rho}{\bar{\rho}} + \Phi
- \frac{\bar{F}_{\varphi } }{2\bar{F}} \frac{1-3 w}{1+w} \delta \varphi \,.
\end{align}

Linearizing equations for the conservation of particle energy-momentum tensor \eqref{EOM-COM-P}, we find the same Eqs. as \eqref{COMP-p-t-u} and \eqref{COMP-p-i-u}.

Linearized equation for the scalar field \eqref{EOM-varphi} gives
\begin{align}\label{KGp-F}
	&\ddot{\delta \varphi}
	+ 3H \dot{\delta \varphi }
	+ \Big( -\frac{1}{a^2} \partial^2 + \bar{V}_{\varphi \varphi } \Big) \delta \varphi
		+ 2 \bar{V}_{\varphi} \Phi
	+ \dot{\bar{\varphi }} (3 \dot{\Psi } - \dot{\Phi }) 
	\nonumber \\
& 	= \frac{\bar{F}_{\varphi } }{2 \bar{F}} \left[ (1 - 3 c_s^2) \delta \rho 
+ 2 \big( \alpha^{(P)} + 3 \psi^{(P)} \big) \right]
\nonumber \\
&+ \frac{1}{2} (1-3 w) \bar{\rho }
\left[ 
2 \frac{\bar{F}_{\varphi}}{\bar{F}}  \Phi 
+
\frac{d}{{\bar N}dt}
\left( \frac{\bar{F}_{\varphi}}{\bar{F}} \right) \delta \varphi 
\right]
	\,,
\end{align}

As it can be seen from the above equation, the minimal coupling and constant scalar background limit $F=\mbox{constant}$ and $\dot{\bar\varphi}=0$ is manifest: there is no source for the gauge-invariant variables $\delta\varphi$ in this limit. Therefore, as expected, there is no scalar memory effect in this limit.


\section{The coupled wave equations on a curved spacetime}\label{app-Green}

In this appendix, our primary aim is to find Green's functions corresponding to Eq. \eqref{Eq-Matrix-ND} starting from the first principle. However, we consider a general setup, which is applicable for other purposes as well. We thus apply our general results in this appendix to the particular case of Eq. \eqref{Eq-Matrix-ND} only in the last subsection.

We start with a set of coupled wave equations propagating on a background spacetime with metric ${g}_{\mu\nu}$ as follows
\begin{align}\label{Eq-Matrix-0}
& {\boldsymbol{\mathcal L}}'. \boldsymbol{\xi}' = - 4\pi \boldsymbol{\mu}' \,; 
& {\boldsymbol{\mathcal L}}' = \boldsymbol{\mathcal G}'^{\mu\nu} { \nabla}_\mu { \nabla}_\nu - 
\boldsymbol{\mathcal N}'^\mu { \nabla}_\mu - \boldsymbol{\mathcal M}' \,,
\end{align}
where $\nabla$ denotes the covariant derivatives compatible with ${g}_{\mu\nu}$, $\boldsymbol{\mathcal G}'^{\mu\nu}$, $\boldsymbol{\mathcal N}'^\mu$ and $\boldsymbol{\mathcal M}'$ are $2\times2$ matrices while $\boldsymbol{\xi}'$ and $\boldsymbol{\mu}'$ are $2\times1$ matrices. The matrices $\boldsymbol{\mathcal G}'^{\mu\nu}$ characterize the lightcone structures of the modes $\boldsymbol{\xi}'$, the matrices $\boldsymbol{\mathcal N}'^\mu$ encode the friction terms, the matrix $\boldsymbol{\mathcal M}'$ is the mass matrix while $\boldsymbol{\mu}'$ is the source matrix. 

Performing field redefinition $\boldsymbol{\xi} = {\bf P}. \boldsymbol{\xi}'$ and defining $ \boldsymbol{\mu} \equiv {\bf P}^\intercal. \boldsymbol{\mu}'$, Eq. \eqref{Eq-Matrix-0} becomes
\begin{align}\label{Eq-Matrix-g}
& \boldsymbol{\mathcal L}. \boldsymbol{\xi} = -4\pi \boldsymbol{\mu} \,; 
& \boldsymbol{\mathcal L} = \boldsymbol{\mathcal G}^{\mu\nu} { \nabla}_\mu { \nabla}_\nu -
\boldsymbol{\mathcal N}^\mu { \nabla}_\mu - \boldsymbol{\mathcal M} \,,
\end{align}
where we have defined
\begin{align}
\boldsymbol{\mathcal G}^{\mu\nu} &\equiv {\bf P}^\intercal.\boldsymbol{\mathcal G}'^{\mu\nu}.{\bf P} \,, 
\label{G-munu} \\
\boldsymbol{\mathcal N}^{\mu} &\equiv {\bf P}^\intercal.\boldsymbol{\mathcal N}'^{\mu}.{\bf P} 
- 2 {\bf P}^\intercal.\boldsymbol{\mathcal G}'^{\mu\nu}.{ \nabla}_\nu{\bf P} \,, 
\label{B-mu} \\
\boldsymbol{\mathcal M} &\equiv {\bf P}^\intercal.\boldsymbol{\mathcal M}'.{\bf P} 
+ {\bf P}^\intercal.\boldsymbol{\mathcal N}'^{\mu}.{ \nabla}_\mu{\bf P} 
- {\bf P}^\intercal.\boldsymbol{\mathcal G}'^{\mu\nu}.{ \nabla}_\mu{ \nabla}_\nu{\bf P} \,.
\label{M}
\end{align}
This transformation will allow us to bring matrix $\boldsymbol{\mathcal G}^{\mu\nu}$ into a simple form by appropriately choosing ${\bf P}$. This is essential when we study the lightcone structures of the modes $\boldsymbol{\xi}$ later. Considering the following component forms
\begin{align}
& \boldsymbol{\xi}
\doteq
\begin{pmatrix}
\xi_1 \\ \xi_2
\end{pmatrix} \,,
& \boldsymbol{\mu}
\doteq
\begin{pmatrix}
\mu_1\\ \mu_2
\end{pmatrix} \,,
\end{align}
Eq. \eqref{Eq-Matrix-g} yields
\begin{align}\label{Eq-xi1}
& \boldsymbol{\mathcal L}_{11} \xi_1 + \boldsymbol{\mathcal L}_{12} \xi_2 = - 4\pi {\mu}_1 \,,\\
\label{Eq-xi2}
& \boldsymbol{\mathcal L}_{21} \xi_1 + \boldsymbol{\mathcal L}_{22} \xi_2 = - 4\pi {\mu}_2 \,,
\end{align}
and the corresponding Green's functions satisfy
\begin{align}
& \boldsymbol{\mathcal L}_{11} {G}_1^{S}(x,x') + \boldsymbol{\mathcal L}_{12} {G}_2^{S}(x,x') 
= -4\pi \delta^{(4)}(x,x') \,, \label{Eq-Greens-xi1}
\\
& \boldsymbol{\mathcal L}_{21} {G}_1^{S}(x,x') + \boldsymbol{\mathcal L}_{22} { G}_2^{S}(x,x') 
= -4\pi \delta^{(4)}(x,x') \,, \label{Eq-Greens-xi2}
\end{align}
where $\delta^{(4)}(x,x')=\delta^{(4)}(x-x')/\sqrt{-{g}}$. The solutions for $\xi_1$ and $\xi_2$ are given by
\begin{align}\label{sol-xi1}
\xi_1(x) &= \int d^4x' \sqrt{-{ g}(x')} {G}^{S}_{1}(x,x') { \mu}_{1}(x') \,,
\\ 
\label{sol-xi2}
\xi_2(x) &= \int d^4x' \sqrt{-{ g}(x')} {G}^{S}_{2}(x,x') { \mu}_{2}(x') \,.
\end{align}

\subsection{Foliation of the spacetime}

The setup in the previous subsection was very general as we did not impose any conditions on the background metric $g_{\mu\nu}$. In this subsection, we simplify the setup by assuming that  the background metric allows for a foliation to time slices. This is a reasonable assumption in the sense that we need a notion of time as far as we are interested in wave equations. For example, this is the case for the cosmological spacetime, spherically symmetric, and black hole solutions. 

Foliating the spacetime region of interest by time slices, the corresponding unit vector normal to the {\it spacelike constant-time hypersurfaces} is given by
\begin{align}
{ n}_{\mu} = -  \frac{\delta^0{}_\mu}{\sqrt{-{ g}^{00}}} \,; \qquad {g}^{\mu\nu}{ n}_{\mu} { n}_{\nu}=-1 \,.
\end{align}
As usual, we define the induced metric on the spatial hypersurfaces as follows 
\begin{equation}\label{h}
{ h}_{\mu\nu}\equiv{ g}_{\mu\nu} + {n}_\mu {n}_\nu \,,
\end{equation}
which satisfies ${ h}_{\mu}{}^\alpha{n}_\alpha=0$ and ${ h}^\mu{}_\alpha{ h}^\alpha{}_\nu={ h}^{\mu}{}_{\nu}$. Based on the existence of the time slices, we assume the following form of the matrices $\boldsymbol{\mathcal G}'^{\mu\nu}$ and $\boldsymbol{\mathcal N}'^{\mu}$. 
\begin{align}\label{metric-g-p}
\boldsymbol{\mathcal G}'^{\mu\nu} &= \boldsymbol{\mathcal C}'\, {h}^{\mu\nu} 
- \boldsymbol{\mathcal K}'\, {n}^\mu {n}^\nu \,,
\\ \label{n-g-p}
\boldsymbol{\mathcal N}'^{\mu} & = \boldsymbol{\Theta}' \, {n}^\mu - \boldsymbol{\kappa}'^\mu \,; 
\hspace{2cm} n_\mu \boldsymbol{\kappa}'^\mu = 0 \,,
\end{align}
where $\boldsymbol{\mathcal C}'$, $\boldsymbol{\mathcal K}'$, $\boldsymbol{\Theta}'$, and $\boldsymbol{\kappa}'^\mu$ are general $2\times2$ matrices and their components are spacetime functions. From \eqref{G-munu} and \eqref{B-mu} we find\footnote{We can also rewrite the metric matrices in the disformal form $\boldsymbol{\mathcal G}^{\mu\nu} = \boldsymbol{\mathcal C}\, {g}^{\mu\nu} + \boldsymbol{\mathcal D}\, {n}^\mu {n}^\nu$ where $\boldsymbol{\mathcal D}\equiv\boldsymbol{\mathcal C}-\boldsymbol{\mathcal K}$. However, the form \eqref{metric-g-p} is more appropriate for our purposes.}
\begin{align}\label{metric-g-g-p}
\boldsymbol{\mathcal G}^{\mu\nu} &= \boldsymbol{\mathcal C} {h}^{\mu\nu}
- \boldsymbol{\mathcal K} {n}^\mu {n}^\nu \,, 
\\ \label{n-g-g-p}
\boldsymbol{\mathcal N}^{\mu} & = \boldsymbol{\Theta} \, {n}^\mu - \boldsymbol{\kappa}^\mu \,; 
\hspace{2cm} n_\mu \boldsymbol{\kappa}^\mu = 0 \,,
\end{align}
where we have defined
\begin{align}
\boldsymbol{\mathcal C} &\equiv {\bf P}^\intercal.\boldsymbol{\mathcal C}'.{\bf P} \,,
\nonumber \\ \nonumber
\boldsymbol{\mathcal K} &\equiv {\bf P}^\intercal.\boldsymbol{\mathcal K}'.{\bf P} \,,
\\ \nonumber
\boldsymbol{\Theta} &\equiv {\bf P}^\intercal.\boldsymbol{\Theta}'.{\bf P} 
- 2 {\bf P}^\intercal.\boldsymbol{\mathcal K}'.n^\nu\nabla_\nu{\bf P} \,,
\\ \nonumber
\boldsymbol{\kappa}^\mu &\equiv {\bf P}^\intercal.\boldsymbol{\kappa}'^\mu.{\bf P}
+ 2 {\bf P}^\intercal.\boldsymbol{\mathcal C}'.h^{\mu\nu}\nabla_\nu{\bf P} \,.
\end{align}
The decomposition \eqref{n-g-g-p} is the most general one for $\boldsymbol{\mathcal N}^{\mu}$ while \eqref{metric-g-g-p} is not the most general one for $\boldsymbol{\mathcal G}^{\mu\nu}$. However, first, this special form is invariant under the general transformation characterized by the matrix ${\bf P}$ as we can see from Eqs. \eqref{metric-g-p} and \eqref{metric-g-g-p}. Second, it is general enough to include many interesting cases, e.g., cosmological, spherically symmetric, and many black hole solutions.

The matrices $\boldsymbol{\mathcal C}$ and $\boldsymbol{\mathcal K}$ characterize the gradient and kinetic terms for the modes $\xi_1$ and $\xi_2$ and, therefore, we assume that they are both positive definite to avoid gradient and ghost instabilities. Moreover, as the tangential and orthogonal parts are independent pieces, two components of ${\bf P}$ are enough to make both $\boldsymbol{\mathcal C}={\bf P}^\intercal.\boldsymbol{\mathcal C}'.{\bf P}$ and $\boldsymbol{\mathcal K}={\bf P}^\intercal.\boldsymbol{\mathcal K}'.{\bf P}$ diagonal. We use the remaining two components of ${\bf P}$ to impose the normalization condition $\boldsymbol{\mathcal K}=\boldsymbol{1}$. Therefore, by appropriately choosing ${\bf P}$, we can always simplify Eq. \eqref{metric-g-g-p} as follows
\begin{align}\label{metric-g}
\boldsymbol{\mathcal G}^{\mu\nu} = \boldsymbol{\mathcal C}\, {h}^{\mu\nu} 
- \boldsymbol{1}\, {n}^\mu {n}^\nu \,,
\end{align}
where now $\boldsymbol{\mathcal C}$ is a diagonal $2\times2$ matrix. Note also that ${ n}_\mu{ n}_\nu\boldsymbol{\mathcal G}^{\mu\nu} = -\boldsymbol{1}$ which shows that choice $\boldsymbol{\mathcal K}=\boldsymbol{1}$, that we have made, corresponds to the normalization conditions for $n^\mu$ with respect to both diagonal components of the metric matrices $\boldsymbol{\mathcal G}^{\mu\nu}$. It is also worth mentioning that we are not interested in the trivial case of $\boldsymbol{\mathcal C}\propto\boldsymbol{1}$, or equivalently $\boldsymbol{\mathcal G}^{\mu\nu}\propto\boldsymbol{1} g^{\mu\nu}$, when both modes propagate with the same speeds. 
The assumption that $\boldsymbol{\mathcal C}$ and $\boldsymbol{\mathcal K}$ are positive definite guaranties the existence of an inverse matrix $\boldsymbol{\mathcal G}^{\mu\alpha}\boldsymbol{\mathcal G}_{\alpha\nu}=\boldsymbol{1} \delta^\mu{}_\nu$ which is given by
\begin{align}\label{metric-g-inv}
\boldsymbol{\mathcal G}_{\mu\nu} = \boldsymbol{\mathcal C}^{-1}\, {h}_{\mu\nu} 
- \boldsymbol{1} \, {n}_\mu {n}_\nu \,.
\end{align}

Now, let us present our results in the component forms which is more useful for some practical purposes. We introduce diagonal components of $\boldsymbol{\mathcal G}^{\mu\nu}$ and $\boldsymbol{\mathcal C}$ as
\begin{align}\label{metric-g-com}
&\boldsymbol{\mathcal G}^{\mu\nu} \equiv \mbox{diag}\left( {\mathcal G}^{\mu\nu}_1,{\mathcal G}^{\mu\nu}_2 \right) \,,
&\boldsymbol{\mathcal C} = {\rm diag}\left( c_1^2,c_2^2 \right) \,.
\end{align}
Eqs. \eqref{metric-g} and \eqref{metric-g-inv} then give
\begin{align}\label{G-metrics-g}
&{\mathcal G}_1^{\mu\nu} = c_1^2 \, { h}^{\mu\nu} - { n}^{\mu} { n}^{\nu} \,,
&{\mathcal G}_2^{\mu\nu} = c_2^2 \, { h}^{\mu\nu} - { n}^{\mu} { n}^{\nu} \,,
\\ \label{G-metrics-g-inv}
&{\mathcal G}_{1\mu\nu} = c_1^{-2} { h}_{\mu\nu} 
- { n}_{\mu} { n}_{\nu} \,,
&{\mathcal G}_{2\mu\nu} = c_2^{-2} { h}_{\mu\nu} 
- { n}_{\mu} { n}_{\nu} \,.
\end{align}
The positivity of $\boldsymbol{\mathcal C}$ together with the condition of excluding the trivial case that both modes propagate at the same speeds, read as
\begin{align}\label{conditions-c_12}
&&{c}_1\neq 0 \,, 
&&{c}_2\neq 0 \,,
&&{c}_1\neq{c}_2 \,.
\end{align}
Using \eqref{n-g-g-p} and \eqref{metric-g} in Eqs.\eqref{Eq-xi1} and \eqref{Eq-xi2} and then using the component forms \eqref{metric-g-com}, we find
\begin{align}
&\left[ \frac{d^2}{ds^2} + \left( c_1^2 \theta + \Theta_{11} \right) \frac{d}{ds} - c_1^2 D^2 - \left( a^\mu+\kappa^\mu_{11} \right) D_\mu + {\mathcal M}_{11} \right] \xi_1 
\nonumber \\ \label{Eq-xi1-app}
&+ \left[ \Theta_{12} \frac{d}{ds} + \kappa^\mu_{12} D_\mu + {\mathcal M}_{12} \right] \xi_2 = 4\pi {\mu}_1 \,,
\\
&\left[ \frac{d^2}{ds^2} + \left( c_2^2 \theta + \Theta_{22} \right) \frac{d}{ds} - c_2^2 D^2 - \left(a^\mu+\kappa^\mu_{22} \right) D_\mu + {\mathcal M}_{22} \right] \xi_2 
\nonumber \\ \label{Eq-xi2-app}
&+ \left[ \Theta_{21} \frac{d}{ds} + \kappa^\mu_{21} D_\mu + {\mathcal M}_{21} \right] \xi_1 = 4\pi {\mu}_2 \,,
\end{align}
where we have defined the parameter $s$ and the expansion scalar\footnote{Indeed, differentiation with respect to the parameter $s$ is nothing but the Lie derivative along the vector $n^\mu$, $d/ds=\pounds_{n}$, which determines the time direction. In the special case when the acceleration vanishes $a_\mu=0$, parameter $s$ becomes the affine parameter of the geodesic equation ${n}^\nu { \nabla}_\nu n_\mu = 0$.}
\begin{align}\label{s-theta}
&& \frac{d}{ds} \equiv { n}^\mu { \nabla}_\mu \,, 
&& \theta \equiv \nabla_\mu n^\mu \,,
\end{align}
and also the spatial derivative and the acceleration
\begin{align}\label{D-a}
&& D_\mu \equiv h_\mu{}^\alpha{ \nabla}_\alpha \,, 
&& a_\mu \equiv { n}^\nu { \nabla}_\nu n_\mu \,.
\end{align}
From Eqs. \eqref{Eq-xi1-app} and \eqref{Eq-xi2-app} we see that $c_1$ and $c_2$ are the sound speeds for the modes $\xi_1$ and $\xi_2$ respectively. This is the advantage of the decomposition based on the time slices.

\subsection{Geodetic intervals and van Vleck-Morette determinants}

As usual, in order to solve the wave equations \eqref{Eq-xi1} and \eqref{Eq-xi2} or equivalently Eqs. \eqref{Eq-xi1-app} and \eqref{Eq-xi2-app}, we define timelike {\it geodetic intervals} $s_1(x,x')$ and $s_2(x,x')$ as follows
\begin{align}\label{s-def}
&{\mathcal G}^{\mu\nu}_{1}{ \nabla}_\mu s_1 {\nabla}_\nu s_1 = -1\,,
&{\mathcal G}^{\mu\nu}_{2}{ \nabla}_\mu s_2 { \nabla}_\nu s_2 = -1 \,.
\end{align}
The geodetic intervals $s_1(x,x')$ and $s_2(x,x')$ are bi-scalars which depend on two different spacetime points $x$ and $x'$. They determine the distance between points $x$ and $x'$ as measured by means of the corresponding metrics along the geodesic joining them. The properties of the bi-scalars are vastly studied in the literature (see for instance Refs. \cite{DeWitt:1960fc,Poisson:2011nh}). We also define other geodetic intervals
\begin{align}\label{s-xi-def}
&\sigma_1(x,x') = - \frac{1}{2} s_1(x,x')^2 \,,
&\sigma_2(x,x') = - \frac{1}{2} s_2(x,x')^2 \,,
\end{align}
which correspond to the geodesic squared distances and satisfy
\begin{align}\label{sigma-xi-def}
&{\mathcal G}^{\mu\nu}_{1}{ \nabla}_\mu \sigma_1 { \nabla}_\nu \sigma_1 
= 2 \sigma_1 \,,
&{\mathcal G}^{\mu\nu}_2{ \nabla}_\mu \sigma_2 { \nabla}_\nu \sigma_2 
= 2 \sigma_2 \,.
\end{align}

It is also useful to express the second covariant derivative of the geodetic intervals $\sigma_1$ and $\sigma_2$ in terms of the so-called van Vleck-Morette determinants
\begin{align}\label{Delta-xi1}
&\Delta_1(x,x') = \frac{D_1(x,x')}{\sqrt{-g(x)}\sqrt{-g(x')}} \,; \\ \nonumber
&D_1(x,x') = - \det[-{ \nabla}_\mu { \nabla}_{\mu'} \sigma_1(x,x')] \,,
\\ \label{Delta-xi2}
&\Delta_2(x,x') = \frac{D_2(x,x')}{\sqrt{-g(x)}\sqrt{-g(x')}} \,; \\ \nonumber
&D_2(x,x') = - \det[-{ \nabla}_\mu { \nabla}_{\mu'} \sigma_2(x,x')] \,,
\end{align}
where $\nabla_{\mu'}$ denotes the covariant derivative with respect to $x'$. Taking covariant derivative of Eqs. \eqref{sigma-xi-def} with respect to $x$ and taking again covariant derivative with respect to $x'$, after manipulating the results in appropriate way \cite{DeWitt:1960fc}, it is straightforward to show that the van Vleck-Morette determinants satisfy the following relations
\begin{align}\label{Delta1}
&\Delta^{-1}_1 { \nabla}_\mu \left( \Delta_1 {\mathcal G}^{\mu\nu}_1 { \nabla}_\nu\sigma_1 \right) = 4 \,, \\
\label{Delta2}
&\Delta^{-1}_2 { \nabla}_\mu \left( \Delta_2 {\mathcal G}^{\mu\nu}_2 { \nabla}_\nu \sigma_2 \right) = 4 \,.
\end{align}
Note that neither metric ${\mathcal G}_{1\mu\nu}$ nor ${\mathcal G}_{2\mu\nu}$ are compatible with the covariant derivative and, therefore, the following nontrivial contributions arise\footnote{Indeed, we could define new covariant derivatives compatible with ${\mathcal G}_{1\mu\nu}$ and ${\mathcal G}_{2\mu\nu}$. In that case, the van Vleck-Morette determinants \eqref{Delta-xi1} and \eqref{Delta-xi2} could be defined completely in terms of ${\mathcal G}_{1\mu\nu}$ and ${\mathcal G}_{2\mu\nu}$ and their determinants. The final result of course does not change and here we found it more convenient to work with the standard covariant derivative compatible with the background metric $g_{\mu\nu}$.}
\begin{align}\label{Dg1}
&{ \nabla}_\mu {\mathcal G}^{\mu\nu}_1 = 2 c_1^2 D^\nu \ln{c_1} + (c_1^2-1) \left( \theta n^\nu + a^\nu \right) \,,
\\\label{Dg2}
&{ \nabla}_\mu {\mathcal G}^{\mu\nu}_2 = 2 c_2^2 D^\nu \ln{c_2} + (c_2^2-1) \left( \theta n^\nu + a^\nu \right) \,.
\end{align}

Let us define the timelike vectors
\begin{align}
t_{1\mu} &\equiv - \nabla_\mu s_1 = \alpha_1\, n_\mu + q_{1\mu} \,,
\nonumber \\ \label{t-def}
t_{2\mu} &\equiv - \nabla_\mu s_2 = \alpha_2\, n_\mu + q_{2\mu} \,,
\end{align}
which are unit vectors ${\mathcal G}^{\mu\nu}_{1} t_{1\mu} t_{1\nu}=-1$ and ${\mathcal G}^{\mu\nu}_{2} t_{2\mu} t_{2\nu}=-1$ by definitions Eqs. \eqref{s-def}. In the above relations, we have decomposed the unit vectors into the orthogonal and tangential pieces which are proportional to $n_\mu$ and the spatial vectors $q_{1,2\mu}$, which satisfies $n^\mu{q}_{1,2\mu}=0$, respectively. The components of $t_{1\mu}$ and $t_{2\mu}$ are spacetime functions which are subject to the normalization conditions
\begin{align}\label{normalizations-ab}
&c_1^2 q_1^2 - \alpha_1^2 = - 1 \,,
&c_2^2 q_2^2 - \alpha_2^2 = - 1 \,,
\end{align}
where $q_{1,2}^2=g^{\mu\nu}q_{1,2\mu}q_{1,2\nu}$. Taking derivative of \eqref{s-xi-def} with respect to the coordinate $x$, we find
\begin{align}
&\nabla_\mu \sigma_1 = s_1\, t_{1\mu} \,,
&\nabla_\mu \sigma_2 = s_2\, t_{2\mu} \,.
\end{align}
Substituting the above result in Eqs. \eqref{Delta1} and \eqref{Delta2} and then using Eqs. \eqref{G-metrics-g} and \eqref{t-def}, after some simplifications, it is straightforward to find the following equations
\begin{align}
& \frac{d}{ds_1} \log(\alpha_1 \Delta_1) 
-  q_1^\mu \left[ \left( 1 - c_1^2 \right) D_\mu \log\Delta_1 + D_\mu\log\alpha_1 \right]
\nonumber \\ \label{Delta1-eq}
& + D_\mu \left( c_1^2 q_1^\mu \right)
= \frac{3}{s_1} - \alpha_1 \theta \,,
\\
&
\frac{d}{ds_2} \log(\alpha_2 \Delta_2) 
- q_2^\mu \left[ \left( 1 - c_2^2 \right) D_\mu \log\Delta_2 + D_\mu\log\alpha_2 \right]
\nonumber \\ \label{Delta2-eq}
&+ D_\mu \left( c_2^2 q_2^\mu \right)
= \frac{3}{s_2} - \alpha_2 \theta \,,
\end{align}
where
\begin{align}
&\frac{d}{ds_1} \equiv t_1^\mu\nabla_\mu = \alpha_1 \frac{d}{ds} + q_1^\mu D_\mu \,,
\nonumber \\ \label{s1-s2}
&\frac{d}{ds_2} \equiv t_2^\mu\nabla_\mu = \alpha_2 \frac{d}{ds} + q_2^\mu D_\mu \,,
\end{align}
are the Lie derivatives along the timelike vectors $t^\mu_1$ and $t^\mu_2$. 

\subsection{Fundamental solutions}

Having geodetic intervals in hand, we consider the following Hadamard representation for the {\it fundamental solutions} of Eqs. \eqref{Eq-xi1} and \eqref{Eq-xi2} or equivalently Eqs. \eqref{Eq-xi1-app} and \eqref{Eq-xi2-app} \cite{DeWitt:1960fc,Garabedian,Friedlander:2010eqa}
\begin{align}\label{G-1}
{ G}_1 
&= \frac{1}{\pi} \left[ \frac{U_1}{\sigma_1} - V_1 \log|\sigma_1| + W_1 + \frac{\hat{U}_2}{\sigma_2} - \hat{V}_2 \log|\sigma_2| \right]
\,,
\\ \label{G-2}
{ G}_2
&= \frac{1}{\pi} \left[ \frac{U_2}{\sigma_2} - V_2 \log|\sigma_2| + {W}_2
+ \frac{\hat{U}_1}{\sigma_1} - \hat{V}_1 \log|\sigma_1| \right]
\,,
\end{align}
where $U_{1,2}$, $V_{1,2}$, $W_{1,2}$, $\hat{U}_{1,2}$, and $\hat{V}_{1,2}$ are bi-scalars which are regular functions of $x$ and $x'$. The bi-scalars $\hat{U}_2$ and $\hat{V}_2$ are introduced to take into account the possibility that $\xi_2$ may induce singularities on the solution of $\xi_1$ through their interaction and the bi-scalars $\hat{U}_1$ and $\hat{V}_1$ are introduced for the similar reason respectively. The bi-scalars $U_1$, $V_1$, and $W_1$ characterize the direct, tail, and regular parts of the mode $\xi_1$ along the lightcone defined by $\sigma_1$ while the bi-scalars $\hat{U}_2$ and $\hat{V}_2$ characterize direct and tail parts of the mode $\xi_1$ along the lightcone defined by $\sigma_2$. 

Substituting the ansatz \eqref{G-1} in Eq. \eqref{Eq-xi1}, the left hand side can be classified into the terms proportional to $\sigma_1^{-2}$, $\sigma_1^{-1}$, $\log|\sigma_1|$, and also $\sigma_2^{-3}$, $\sigma_2^{-2}$, $\sigma_2^{-1}$, $\log|\sigma_2|$. Note that the term which is apparently proportional to $\sigma_1^{-3}$ gives a contribution of the order of $\sigma_1^{-2}$ after substituting the result \eqref{sigma-xi-def}. Note also that the terms which are singular in $\sigma_2$ appear due to the interaction between $\xi_1$ and $\xi_2$. Similarly, substituting ansatz \eqref{G-2} in Eq. \eqref{Eq-xi2}, the left hand side can be classified into the terms proportional to $\sigma_2^{-2}$, $\sigma_2^{-1}$, $\log|\sigma_2|$, and also $\sigma_1^{-3}$, $\sigma_1^{-2}$, $\sigma_1^{-1}$, $\log|\sigma_1|$. Demanding that the coefficients of the terms with the highest degree of singularity, i.e. $\sigma_2^{-3}$ and $\sigma_1^{-3}$, vanish, we immediately conclude
\begin{align}\label{Uhat-zero-g}
&\hat{U}_1 = 0 \,, 
&\hat{U}_2 = 0 \,.
\end{align}
Going to the next order and demanding that the coefficients of the terms which are proportional to $\sigma_2^{-2}$ and $\sigma_1^{-2}$ in Eqs. \eqref{Eq-xi1} and \eqref{Eq-xi2} vanish, we find
\begin{align}
\hat{V}_1 &= - \left(\frac{c_1^2}{c_2^2-c_1^2}\right) \left(\frac{d\sigma_1}{ds}\right)^{-1} \Theta_{21} \, U_1 \,, 
\nonumber \\ \label{Vhat-g}
\hat{V}_2 &= \left(\frac{c_2^2}{c_2^2-c_1^2}\right) \left(\frac{d\sigma_2}{ds}\right)^{-1}\Theta_{12} \, U_2 \,,
\end{align}
where we have used \eqref{Uhat-zero-g}. Note that the denominators become singular when $c_1=c_2$ which is the case that we have excluded in our analysis from the beginning in Eq. \eqref{conditions-c_12}. The results \eqref{Uhat-zero-g} and \eqref{Vhat-g} show that the mode $\xi_2$ does not contribute to the direct part of the solution ${ G}_1$ since $\hat{U}_2=0$. However, the direct part of $\xi_2$ contributes to the tail of the mode $\xi_1$ since $\hat{V}_2 \propto {U}_2$.

Now, we look at the coefficients of the terms which are proportional to $\sigma_1^{-2}$ and $\sigma_2^{-2}$ in Eqs. \eqref{Eq-xi1} and \eqref{Eq-xi2} and by demanding that they vanish, we find the following equations for the direct parts 
\begin{align}
&\Big\{
{\mathcal G}^{\mu\nu}_1 \left[ 2 { \nabla}_\mu {U}_1 
- ( { \nabla}_\mu \log\Delta_1 ) U_1 \right] 
\nonumber \\ \label{Eq-U-xi1}
&- \left( \Theta_{11} { n}^\nu - \kappa_{11}^\nu + \nabla_\mu {\mathcal G}_1^{\mu\nu} \right) U_{1} 
\Big\} { \nabla}_\nu \sigma_1 = 0 \,,
\\
&\Big\{
{\mathcal G}^{\mu\nu}_2 \left[ 2 { \nabla}_\mu {U}_2 
- ( { \nabla}_\mu \log\Delta_2 ) U_2 \right] 
\nonumber \\ \label{Eq-U-xi2}
&- \left( \Theta_{22} { n}^\nu - \kappa_{22}^\nu + \nabla_\mu {\mathcal G}_2^{\mu\nu} \right) U_2 
\Big\} { \nabla}_\nu \sigma_2 = 0 \,,
\end{align}
which are subject to the initial conditions
\begin{align}\label{U-xi-lim}
&\lim\limits_{x\to{x'}}U_{1}(x,x') = 1 \,,
&\lim\limits_{x\to{x'}}U_{2}(x,x') = 1
\,.
\end{align}
In principle, we can always solve the first order equations \eqref{Eq-U-xi1} and \eqref{Eq-U-xi2} to find ${U}_1$ and ${U}_2$ in terms of the van Vleck-Morette determinants $\Delta_1$ and $\Delta_2$. However, we can further simplify these results. Using the explicit forms of the metrics \eqref{G-metrics-g} and also the results \eqref{Dg1} and \eqref{Dg2} in Eqs. \eqref{Eq-U-xi1} and \eqref{Eq-U-xi2}, and then contracting the results with $n_\mu$, we find
\begin{align}\label{Eq-U-xi1-f}
\frac{d}{ds} \left[ \log\left(\frac{{U}_1}{\sqrt{\Delta_1}}\right) \right] 
& = - \frac{1}{2} \left[ \Theta_{11} - \left( 1 - c_1^2 \right) \theta \right] \,,
\\ \label{Eq-U-xi2-f}
\frac{d}{ds} \left[ \log\left(\frac{{U}_2}{\sqrt{\Delta_2}}\right) \right] 
& = - \frac{1}{2} \left[ \Theta_{22} - \left( 1 - c_2^2 \right) \theta \right] \,,
\end{align}
and contracting with $h_{\mu\nu}$, we find
\begin{align}\label{Eq-U-xi1-f-spatial}
c_1^2 D_\mu \left[ \log\left(\frac{{U}_1}{c_1\sqrt{\Delta_1}}\right) \right] 
& = - \frac{1}{2} \left[ \kappa_{\mu 11} + \left( 1 - c_1^2 \right) a_\mu \right] \,,
\\ \label{Eq-U-xi2-f-spatial}
c_2^2 D_\mu \left[ \log\left(\frac{{U}_2}{c_2\sqrt{\Delta_2}}\right) \right] 
& = - \frac{1}{2} \left[ \kappa_{\mu 22} + \left( 1 - c_2^2 \right) a_\mu \right] \,.
\end{align}
The set of equations \eqref{Eq-U-xi1-f}-\eqref{Eq-U-xi2-f-spatial} admits a solution at least locally, provided that they satisfy integrability conditions.

\subsection{Green's functions}

Having the fundamental solutions \eqref{G-1} and \eqref{G-2} in hand, we can easily find the Green's functions by going to the complex plane and performing the so-called $i\epsilon$ prescription. In this regard, we introduce the Feynman propagators as follows
\begin{align}\label{GF-xi1}
{ G}^F_1 
&= \frac{1}{\pi} \left[ \frac{U_1}{\sigma_1+i\epsilon} - V_1 \log|\sigma_1+i\epsilon| + W_1 - \hat{V}_2 \log|\sigma_2+i\epsilon| \right]
\,,
\\ \label{GF-xi2}
{ G}^F_2
&= \frac{1}{\pi} \left[ \frac{U_2}{\sigma_2+i\epsilon} - V_2 \log|\sigma_2+i\epsilon| + {W}_2
- \hat{V}_1 \log|\sigma_1+i\epsilon| \right]
\,,
\end{align}
where we have used the result \eqref{Uhat-zero-g}. Separating the Feynman propagators into the real and imaginary parts as 
\begin{align}
&{ G}^F_1 = {G}_1 - i { G}^S_1 \,, 
&{ G}^F_2 = { G}_2 - i { G}^S_2 \,,
\end{align}
and using the identities 
\begin{align}
&\frac{1}{\sigma+i\epsilon} = {\mathcal P}\left(\frac{1}{\sigma}\right)-i\pi\delta(\sigma) \,,
\nonumber \\ \nonumber
&\log(\sigma+i\epsilon)=\log|\sigma|+i\pi\Theta(-\sigma) \,,
\end{align}
where ${\mathcal P}$ denotes principal value, we find the following expressions for the Green's functions
\begin{align}
{ G}^{S}_1(x,x') 
&= U_{1}(x,x') \delta(\sigma_1) + V_{1}(x,x') \Theta(-\sigma_1) 
\nonumber \\ \label{Greens-xi1-0}
&+ \hat{V}_2(x,x') \Theta(-\sigma_2)
\,, \\
{ G}^{S}_2(x,x') 
&= U_2(x,x') \delta(\sigma_2) + V_2(x,x') \Theta(-\sigma_2) 
\nonumber \\ \label{Greens-xi2-0}
&+ \hat{V}_{1}(x,x') \Theta(-\sigma_1)
\,.
\end{align}

Eqs. \eqref{Greens-xi1-0} and \eqref{Greens-xi2-0} are our final results for the Green's functions of the two modes $\xi_1$ and $\xi_2$ which satisfy the wave equations \eqref{Eq-xi1} and \eqref{Eq-xi2}. Having the Green's functions in hand, one can find solutions of $\xi_1$ and $\xi_2$ for any sources $\mu_1$ and $\mu_2$ through the Eqs. \eqref{sol-xi1} and \eqref{sol-xi2} as usual. We have found explicit equations for the direct parts which are generally given by Eqs. \eqref{Eq-U-xi1-f}, \eqref{Eq-U-xi2-f} and \eqref{Eq-U-xi1-f-spatial}, \eqref{Eq-U-xi2-f-spatial}. It is also straightforward to find the tail and regular parts following the same strategy that we have found Eqs. \eqref{Uhat-zero-g}, \eqref{Vhat-g}, \eqref{Eq-U-xi1} and \eqref{Eq-U-xi2}. We, however, do not need them for our purpose of studying the memory effect and we do not present them here.

\subsection{Application: scalar perturbations in scalar-tensor theory}

Let us now turn back to the system studied in the main text which is a particular subset of the general setup investigated in the previous subsections. Eq. \eqref{Eq-Matrix-ND} can be rewritten in the form of Eq. \eqref{Eq-Matrix-g} through the identifications
\begin{align}
&\xi_1=\zeta \,, 
&\xi_2=\delta{Q} \,,
\end{align}
and considering the subset 
\begin{align}\label{matrices-model}
&&\boldsymbol{\Theta} = {\bf N} - 3H{\bf C} \,,
&&\boldsymbol{\mathcal C} = {\bf C} \,,
&&\boldsymbol{\mathcal M}={\bf M} \,,
&&\boldsymbol{\mu}=\boldsymbol{\mu}^{(P)} \,,
\end{align}
where the explicit forms of the matrices ${\bf N}$, ${\bf C}$, ${\bf M}$, and $\boldsymbol{\mu}^{(P)}$ are given in Eqs. \eqref{K-components}, \eqref{G-components}, \eqref{M-components}, and \eqref{mu-components}. Using \eqref{G-components} in \eqref{matrices-model} and then substituting in \eqref{metric-g}, we find 
\begin{align}\label{G-metrics}
&{\mathcal G}^{\mu\nu}_\zeta = {\bar h}^{\mu\nu} - \bar{n}^\mu \bar{n}^\nu = {\bar g}^{\mu\nu} \,,
&{\mathcal G}^{\mu\nu}_{\delta{Q}} = c_s^2 \bar{h}^{\mu\nu} - \bar{n}^\mu \bar{n}^\nu \,,
\end{align}
where the background metric $g_{\mu\nu}={\bar g}_{\mu\nu}$ is considered. We see that metrics become diagonal when we work with $\delta{Q}$ instead of $\delta\chi$ in Eq. \eqref{Eq-Matrix-ND}. Had we worked with $\delta\chi$, we had to start from \eqref{Eq-Matrix-0} and then finding an appropriate form for matrix ${\bf P}$ which makes $\boldsymbol{\mathcal G}'^{\mu\nu}$ diagonal through Eq. \eqref{G-munu}. 

From Eqs. \eqref{Greens-xi1-0} and \eqref{Greens-xi2-0} we find the following expressions for the corresponding retarded Green's functions
\begin{align}
{\hat G}^{\rm ret}_\zeta(x,x') 
&= \Big[ U_{\zeta}(x,x') \delta(\sigma_\zeta) + V_{\zeta}(x,x') \Theta(-\sigma_\zeta) 
\nonumber \\ \label{Greens-zeta}
&+ \hat{V}_{\delta{Q}}(x,x') \Theta(-\sigma_{\delta{Q}}) \Big] \Theta(t-t')
\,, \\
{\hat G}^{\rm ret}_{\delta{Q}}(x,x') 
&= \Big[ U_{\delta{Q}}(x,x') \delta(\sigma_{\delta{Q}}) + V_{\delta{Q}}(x,x') \Theta(-\sigma_{\delta{Q}}) 
\nonumber \\ \label{Greens-Q}
&+ \hat{V}_{\zeta}(x,x') \Theta(-\sigma_\zeta) \Big] \Theta(t-t')
\,,
\end{align}
where the geodetic intervals satisfy
\begin{align}\label{sigma-zeta-Q-def}
&{\mathcal G}^{\mu\nu}_{\zeta}{\bar \nabla}_\mu \sigma_\zeta {\bar \nabla}_\nu \sigma_\zeta 
= 2 \sigma_\zeta \,,
&{\mathcal G}^{\mu\nu}_{\delta{Q}}{\bar \nabla}_\mu \sigma_{\delta{Q}} {\bar \nabla}_\nu \sigma_{\delta{Q}} 
= 2 \sigma_{\delta{Q}} \,,
\end{align}
in which bars indicate that the covariant derivatives are defined in the spirit of background metric ${\bar g}_{\mu\nu}$. 

Finding an explicit solution for the direct part of $\zeta$ with $c_1=1$ and $a_\mu=0$ is easy. In this case, Eq. \eqref{Delta1-eq} significantly simplifies and the van Vleck-Morette determinant can be written as follows \cite{Visser:1992pz,Ivanov:2018cyw}
\begin{align}\label{van-Vleck-g}
&\Delta_1(x,x') = \exp\left[ \int_{s(x')}^{s(x)} \left( \frac{3}{s}-\theta \right) ds \right] \,,
\end{align}
where the initial condition \eqref{U-xi-lim} is imposed. Integrating Eq. \eqref{Eq-U-xi1-f} from $x'$ to $x$, after using \eqref{van-Vleck-g}, we find
\begin{align}\label{U-1-final}
U_1(x,x') & = s(x,x')^\frac{3}{2} \exp\left[
- \frac{1}{2} \int_{s(x')}^{s(x)} \left(\Theta_{11}+\theta\right) ds
\right]
\,.
\end{align}

The explicit form of the spatially flat FLRW background metric is given by Eq. \eqref{BG-gravity}. Working with conformal time $\eta$ defined in Eq. \eqref{BG-gravity-C}, we have $ds = a d\eta$ and $\theta(\eta) =3a^{-2}da/d\eta$. The direct part then can be obtained from Eq. \eqref{U-1-final} as follows
\begin{align}
&U_{\zeta}(\eta,\eta') = \frac{A_\zeta (\eta')}{A_\zeta(\eta)} \,,
\nonumber \\ \label{U-final}
&A_\zeta(\eta) \equiv a(\eta) \exp\left[\frac{1}{2}\int_\eta 
\left( a N_{11} - \frac{3}{a} \frac{da}{d{\bar \eta}} \right)d{\bar \eta}\right] \,,
\end{align}
which coincides with the result \eqref{U} that is found by another approach.

\bibliography{ref}

\end{document}